\documentclass[%
 amsmath,amssymb,
 aps, physrev,
]{revtex4-2}
\usepackage[colorlinks]{hyperref}
\usepackage{multirow}
\usepackage{wasysym}
\usepackage{graphicx}
\usepackage{dcolumn}
\usepackage{bm}

\begin{document}

\preprint{APS/123-QED}

\title{\boldmath A new dark matter direct search based on archaeological Pb}

\author{D.~Alloni}
\affiliation{Laboratorio Energia Nucleare Applicata, Via Aselli 41, I-27100 Pavia, Italy}

\author{G.~Benato}
\affiliation{Gran Sasso Science Institute, Viale F. Crispi 7, I-67100 L’Aquila, Italy}
\affiliation{INFN Laboratori Nazionali del Gran Sasso, Via G. Acitelli 22, I-67100 Assergi, Italy}

\author{P.~Carniti}
\affiliation{Dipartimento di Fisica, Università di Milano - Bicocca, Piazza della Scienza 3, I-20126 Milano, Italy}
\affiliation{INFN Sezione di Milano - Bicocca, Piazza della Scienza 3, I-20126 Milano, Italy}

\author{M.~Cataldo}
\affiliation{Dipartimento di Fisica, Università di Milano - Bicocca, Piazza della Scienza 3, I-20126 Milano, Italy}
\affiliation{INFN Sezione di Milano - Bicocca, Piazza della Scienza 3, I-20126 Milano, Italy}

\author{L.~Chen}
\affiliation{Shanghai Institute of Ceramics, CAS, 1295 Dingxi Road, Shanghai 200050, P.R. China}

\author{M.~Clemenza}
\affiliation{Dipartimento di Fisica, Università di Milano - Bicocca, Piazza della Scienza 3, I-20126 Milano, Italy}
\affiliation{INFN Sezione di Milano - Bicocca, Piazza della Scienza 3, I-20126 Milano, Italy}

\author{M.~Consonni}
\affiliation{Dipartimento di Fisica, Università di Milano - Bicocca, Piazza della Scienza 3, I-20126 Milano, Italy}
\affiliation{INFN Sezione di Milano - Bicocca, Piazza della Scienza 3, I-20126 Milano, Italy}

\author{G.~Croci}
\affiliation{Dipartimento di Fisica, Università di Milano - Bicocca, Piazza della Scienza 3, I-20126 Milano, Italy}
\affiliation{INFN Sezione di Milano - Bicocca, Piazza della Scienza 3, I-20126 Milano, Italy}

\author{I.~Dafinei}
\affiliation{INFN Sezione di Roma, P.le Aldo Moro 2, I-00185 Roma, Italy}

\author{F.A.~Danevich}
\affiliation{Institute for Nuclear Research of NASU, 03028 Kyiv, Ukraine}
\affiliation{INFN Sezione di Roma Tor Vergata, I-00133 Rome, Italy}

\author{D.~Di~Martino}
\affiliation{Dipartimento di Fisica, Università di Milano - Bicocca, Piazza della Scienza 3, I-20126 Milano, Italy}
\affiliation{INFN Sezione di Milano - Bicocca, Piazza della Scienza 3, I-20126 Milano, Italy}

\author{E.~Di~Stefano}
\affiliation{INFN Sezione di Milano - Bicocca, Piazza della Scienza 3, I-20126 Milano, Italy}
\affiliation{DISAT, Università di Milano - Bicocca, Piazza della Scienza 1, I-20126 Milano, Italy}

\author{N.~Ferreiro Iachellini}
\email{Contact author: nahuel.ferreiroiachellini@ubnimib.it}
\affiliation{Dipartimento di Fisica, Università di Milano - Bicocca, Piazza della Scienza 3, I-20126 Milano, Italy}
\affiliation{INFN Sezione di Milano - Bicocca, Piazza della Scienza 3, I-20126 Milano, Italy}

\author{F.~Ferroni}
\affiliation{Gran Sasso Science Institute, Viale F. Crispi 7, I-67100 L’Aquila, Italy}
\affiliation{INFN Sezione di Roma, P.le Aldo Moro 2, I-00185 Roma, Italy}

\author{F.~Filippini}
\affiliation{INFN Sezione di Milano - Bicocca, Piazza della Scienza 3, I-20126 Milano, Italy}
\affiliation{DISAT, Università di Milano - Bicocca, Piazza della Scienza 1, I-20126 Milano, Italy}

\author{S.~Ghislandi}
\email{Contact author: stefano.ghislandi@gssi.it}
\affiliation{Gran Sasso Science Institute, Viale F. Crispi 7, I-67100 L’Aquila, Italy}
\affiliation{INFN Laboratori Nazionali del Gran Sasso, Via G. Acitelli 22, I-67100 Assergi, Italy}

\author{A.~Giachero}
\affiliation{Dipartimento di Fisica, Università di Milano - Bicocca, Piazza della Scienza 3, I-20126 Milano, Italy}
\affiliation{INFN Sezione di Milano - Bicocca, Piazza della Scienza 3, I-20126 Milano, Italy}

\author{L.~Gironi}
\affiliation{Dipartimento di Fisica, Università di Milano - Bicocca, Piazza della Scienza 3, I-20126 Milano, Italy}
\affiliation{INFN Sezione di Milano - Bicocca, Piazza della Scienza 3, I-20126 Milano, Italy}

\author{P.~Gorla}
\affiliation{INFN Laboratori Nazionali del Gran Sasso, Via G. Acitelli 22, I-67100 Assergi, Italy}

\author{C.~Gotti}
\affiliation{Dipartimento di Fisica, Università di Milano - Bicocca, Piazza della Scienza 3, I-20126 Milano, Italy}
\affiliation{INFN Sezione di Milano - Bicocca, Piazza della Scienza 3, I-20126 Milano, Italy}

\author{D.L.~Helis}
\affiliation{INFN Laboratori Nazionali del Gran Sasso, Via G. Acitelli 22, I-67100 Assergi, Italy}

\author{D.V.~Kasperovych}
\affiliation{Institute for Nuclear Research of NASU, 03028 Kyiv, Ukraine}

\author{V.V.~Kobychev}
\affiliation{Institute for Nuclear Research of NASU, 03028 Kyiv, Ukraine}

\author{G.~Marcucci}
\affiliation{Dipartimento di Fisica, Università di Milano - Bicocca, Piazza della Scienza 3, I-20126 Milano, Italy}
\affiliation{INFN Sezione di Milano - Bicocca, Piazza della Scienza 3, I-20126 Milano, Italy}

\author{A.~Melchiorre}
\affiliation{INFN Laboratori Nazionali del Gran Sasso, Via G. Acitelli 22, I-67100 Assergi, Italy}
\affiliation{Dipartimento di Scienze Fisiche e Chimiche, Università degli Studi dell'Aquila, I-67100 L'Aquila, Italy}

\author{A.~Menegolli}
\affiliation{Dipartimento di Fisica, Università di Pavia, Via Bassi 6, I-27100 Pavia, Italy}
\affiliation{INFN Sezione di Pavia, Via Bassi 6, I-27100 Pavia, Italy}

\author{S.~Nisi}
\affiliation{INFN Laboratori Nazionali del Gran Sasso, Via G. Acitelli 22, I-67100 Assergi, Italy}

\author{M.~Musa}
\affiliation{Dipartimento di Scienze della Terra e dell'Ambiente, Università di Pavia, Via Ferrata 7, I-27100 Pavia, Italy}

\author{L.~Pagnanini}
\affiliation{Gran Sasso Science Institute, Viale F. Crispi 7, I-67100 L’Aquila, Italy}
\affiliation{INFN Laboratori Nazionali del Gran Sasso, Via G. Acitelli 22, I-67100 Assergi, Italy}

\author{L.~Pattavina}
\affiliation{Dipartimento di Fisica, Università di Milano - Bicocca, Piazza della Scienza 3, I-20126 Milano, Italy}
\affiliation{INFN Sezione di Milano - Bicocca, Piazza della Scienza 3, I-20126 Milano, Italy}

\author{G.~Pessina}
\affiliation{INFN Sezione di Milano - Bicocca, Piazza della Scienza 3, I-20126 Milano, Italy}

\author{S.~Pirro}
\affiliation{INFN Laboratori Nazionali del Gran Sasso, Via G. Acitelli 22, I-67100 Assergi, Italy}

\author{O.G.~Polischuk}
\affiliation{Institute for Nuclear Research of NASU, 03028 Kyiv, Ukraine}

\author{S.~Pozzi}
\affiliation{Dipartimento di Fisica, Università di Milano - Bicocca, Piazza della Scienza 3, I-20126 Milano, Italy}
\affiliation{INFN Sezione di Milano - Bicocca, Piazza della Scienza 3, I-20126 Milano, Italy}

\author{M.C.~Prata}
\affiliation{INFN Sezione di Pavia, Via Bassi 6, I-27100 Pavia, Italy}

\author{A.~Puiu}
\affiliation{INFN Laboratori Nazionali del Gran Sasso, Via G. Acitelli 22, I-67100 Assergi, Italy}

\author{S.~Quitadamo}
\affiliation{Gran Sasso Science Institute, Viale F. Crispi 7, I-67100 L’Aquila, Italy}
\affiliation{INFN Laboratori Nazionali del Gran Sasso, Via G. Acitelli 22, I-67100 Assergi, Italy}

\author{M.P.~Riccardi}
\affiliation{Dipartimento di Scienze della Terra e dell'Ambiente, Università di Pavia, Via Ferrata 7, I-27100 Pavia, Italy}

\author{M.~Rossella}
\affiliation{INFN Sezione di Pavia, Via Bassi 6, I-27100 Pavia, Italy}

\author{R.~Rossini}
\affiliation{Dipartimento di Fisica, Università di Pavia, Via Bassi 6, I-27100 Pavia, Italy}
\affiliation{INFN Sezione di Pavia, Via Bassi 6, I-27100 Pavia, Italy}

\author{F.~Saliu}
\affiliation{INFN Sezione di Milano - Bicocca, Piazza della Scienza 3, I-20126 Milano, Italy}
\affiliation{DISAT, Università di Milano - Bicocca, Piazza della Scienza 1, I-20126 Milano, Italy}

\author{A.~Salvini}
\affiliation{Laboratorio Energia Nucleare Applicata, Via Aselli 41, I-27100 Pavia, Italy}

\author{A.P.~Scherban}
\affiliation{National Science Center 'Kharkiv Institute of Physics and Technology', 61108 Kharkiv, Ukraine}

\author{D.A.~Solopikhin}
\affiliation{National Science Center 'Kharkiv Institute of Physics and Technology', 61108 Kharkiv, Ukraine}

\author{V.I.~Tretyak}
\affiliation{INFN Laboratori Nazionali del Gran Sasso, Via G. Acitelli 22, I-67100 Assergi, Italy}
\affiliation{Institute for Nuclear Research of NASU, 03028 Kyiv, Ukraine}

\author{D.~Trotta}
\affiliation{Dipartimento di Fisica, Università di Milano - Bicocca, Piazza della Scienza 3, I-20126 Milano, Italy}
\affiliation{INFN Sezione di Milano - Bicocca, Piazza della Scienza 3, I-20126 Milano, Italy}

\author{H.~Yuan}
\affiliation{Shanghai Institute of Ceramics, CAS, 1295 Dingxi Road, Shanghai 200050, P.R. China}

\collaboration{RES-NOVA Collaboration}
\email{Contact collaboration: res-nova@unimib.it}

\date{\today}

\begin{abstract}
The RES-NOVA project is an experimental initiative aimed at detecting neutrinos from the next galactic supernova using PbWO$_{4}$ cryogenic detectors, operated at low temperatures in a low-background environment. By utilizing archaeological lead (Pb) as the target material, RES-NOVA leverages its high radiopurity, large nuclear mass, and the natural abundance of $^{207}$Pb, making it well-suited for exploring both spin-independent and spin-dependent Dark Matter (DM) interactions via nuclear scattering.

This work presents a background model developed for the RES-NOVA technology demonstrator and evaluates its implications for Dark Matter detection. Detailed calculations of nuclear matrix elements, combined with the unique properties of archaeological Pb, demonstrate RES-NOVA's potential as a complementary tool to existing direct detection experiments for studying Dark Matter interactions. The experiment will conduct DM searches over a broad mass range spanning 4 orders of magnitude, from sub-GeV/$c^2$ to TeV/$c^2$. In the most optimistic scenario of 1~y of data taking, RES-NOVA is expected to probe DM-nucleon cross-sections down to 1$\times 10^{-43}$~cm$^2$ and 2$\times 10^{-46}$~cm$^2$ for candidates with masses of 2~GeV/$c^2$ and 20~GeV/$c^2$, respectively.
\end{abstract}

\maketitle


\section{Introduction}
\label{intro}
The quest to unravel the mysteries of Dark Matter (DM) constitutes one of the most compelling endeavors in modern astrophysics and particle physics~\cite{Billard_2022}. Despite being one of the dominant constituents of the universe, accounting for approximately 85\%~\cite{ParticleDataGroup} of its total mass, DM eludes direct detection and understanding. Its presence is inferred from gravitational effects on visible matter, radiation, and the large-scale structure of the universe~\cite{Cirelli:2024ssz}. Among the myriad of experimental efforts aimed at detecting DM, the search for innovative and more sensitive observational methods remains paramount.
In this context, the RES-NOVA project emerges as a new initiative~\cite{Pattavina:2020cqc}. Originally designed to detect neutrinos from core-collapse supernovae (SN) via coherent elastic neutrino-nucleus scattering (CE$\nu$NS), \mbox{RES-NOVA} utilizes an array of cryogenic detectors~\cite{Pirro:2017ecr} made of archaeological Pb~\cite{Pattavina:2019pxw}. This choice of material is based on its ultra-high radiopurity and the significant CE$\nu$NS cross-section on Pb~\cite{Drukier:1983gj}, making it an exemplary candidate for neutrino astronomy and astrophysical phenomena investigation. In literature, it was also proposed to exploit RES-NOVA sensitivity to all-flavors SN neutrinos: to dramatically improve the sensitivity to the all-flavors detection of the Diffuse Supernova Neutrino Background~\cite{Suliga_DSNB}, to constraint nuclear physics parameters, such as Pb neutron skin~\cite{SN_skin}, but also to search for physics beyond the Standard Model~\cite{Sen:2024fxa}.

The innovative approach adopted by RES-NOVA, combined with its sensitivity, also presents an intriguing possibility: to extend the reach of CE$\nu$NS-based SN neutrino detectors to directly observe DM particles from our galactic halo. Theoretical models, like the one presented in Refs.~\cite{Catena:2013pka,SILVEIRA1985136, Kanemura:2010sh} just to name a few, suggest that DM, in the form of Weakly Interacting Massive Particles (WIMPs and WIMP-like) or other candidate particles~\cite{ARBEY2021103865, NUSSINOV198555}, could interact with ordinary matter in a manner detectable by the RES-NOVA setup. Such interactions, albeit expected to be rare and challenging to discern from background noise, could be within the detection capabilities of RES-NOVA.

The requirements for the observation of low-energy neutrino signals from the cosmos via CE$\nu$NS have strong similarities to the ones of experiments currently searching for DM via its direct interaction with terrestrial detectors~\cite{Billard}. This type of investigation requires the operation of experimental set-up in low-background conditions, given the low interaction cross-section of DM with conventional matter, and low-energy threshold for the detection of the low-energy nuclear recoil induced by the DM interaction~\cite{Sadoulet}.

This paper aims to explore the potential of the RES-NOVA detector in the direct detection of DM, focusing on its physics reach. We discuss the key aspects of the RES-NOVA detector and how its design and operational principles could be tailored or enhanced to probe the DM populating our galactic halo. We also present a preliminary background model that is needed as input to constraint the fundamental properties of DM, namely its mass and the interaction cross-section with the target nuclei. We will consider both elastic scattering on nuclei with spin-dependent (SD) and spin-indepenent (SI) interactions, taking advantage of the isotopic composition of Pb. Finally, some conclusion and outlook in the overall DM framework will be presented.

\section{Astrophysical neutrinos signature via CE$\nu$NS in RES-NOVA}
\label{sec:1}

\begin{figure}[]
\centering
\includegraphics[width=.6\textwidth]{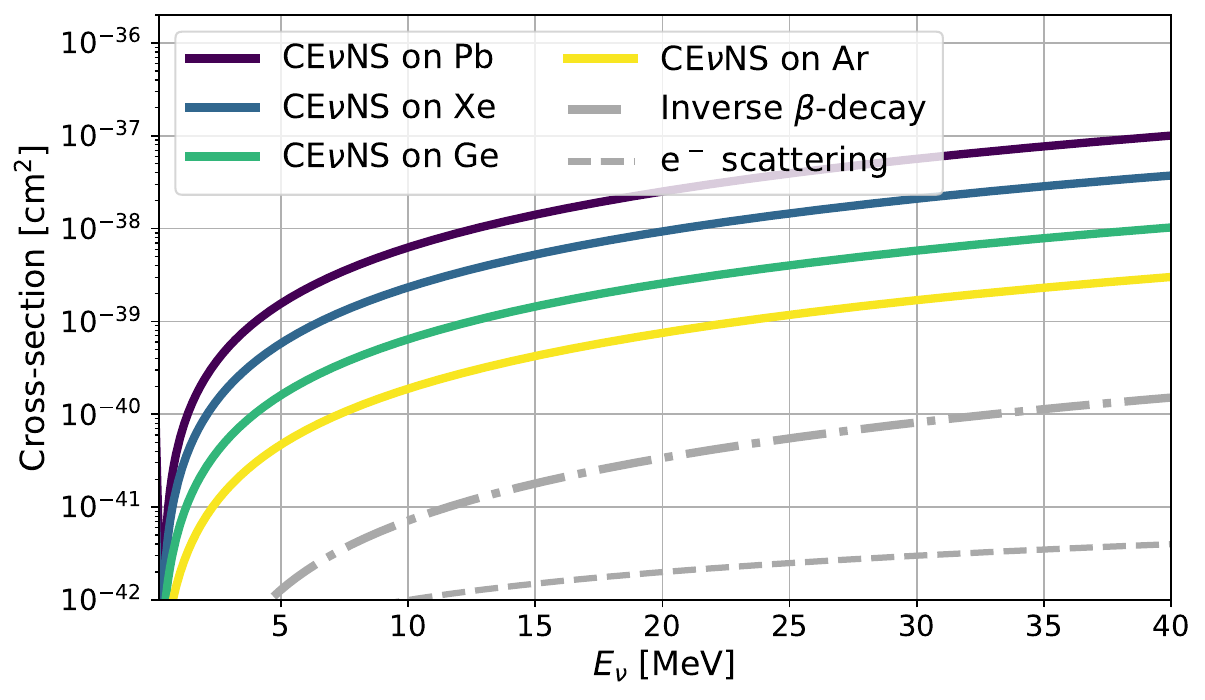}
\caption{Coherent elastic neutrino-nucleus cross-sections (solid lines) as a function of the interacting neutrino energy for different target materials. Inverse-beta decay and neutrino-electron elastic scattering cross-sections are shown as dashed lines. \label{fig:sigma}}
\end{figure}

CE$\nu$NS is a neutral current process~\cite{Freedman:1973yd}, equally sensitive to all neutrino flavors, featuring an extremely high cross-section when compared with other widely used detection channels like neutrino-electron elastic-scattering or inverse-beta decay~\cite{Scholberg:2012id}, see Fig.~\ref{fig:sigma}. 
The interaction cross-section can be derived by basic principles of the Standard Model, and for a spin zero interactions~\cite{Freedman:1973yd} it follows:
\begin{eqnarray}
\label{eq:xsec}
\frac{d\sigma}{d E_R} = \frac{G^2_F m_N}{8 \pi (\hbar c )^4} \left[(4\sin^2 \theta_W -1 ) Z + N \right]^2 
\cdot \left(2- \frac{E_R m_N}{E_{\nu}^2} \right) \cdot |F(q)|^2 ,
\end{eqnarray}
where $G_F$ is the Fermi coupling constant, $Z$ and $N$ the atomic and neutron numbers of the target nucleus, while $m_N$ its mass,  $\theta_W$ the Weinberg angle, $E_{\nu}$ the neutrino energy and $E_R$ the recoil energy of the target. Finally, $F(q)$, is the elastic nuclear form factor at momentum transfer $q=\sqrt{2E_R m_N}$, also known as \textit{coherent factor}. This equation shows that the interaction cross-section approximately scales with $N^2$, with negligible contribution from the proton content of the nucleus~\footnote{($4 \sin^2 \theta_W -1) \approx 0.05$.}. As a consequence, target nuclei with large neutron number lead to higher interaction cross-section.

In this context, Pb is an ideal target material, and for this reason RES-NOVA aims to exploit the advantages of CE$\nu$NS by using cryogenic detectors made from archaeological Pb, for detecting the tiny recoils of nuclei resulting from CE$\nu$NS interactions, which are on the order of just a few keV for SN neutrino interactions. The expected signature is an exponentially increasing rate of events at low energies. 
The use of Pb of archaeological origin is driven by its intrinsic ultra-low content of radioactive impurities~\cite{Pattavina:2019pxw}, like the ancient Roman one~\cite{Nosengo}. In fact, its ancient age (over 2000 years) and prolonged underwater storage have provided two key advantages: a significant reduction in residual radioactivity, particularly from  $^{210}$Pb, and a reduced  cosmogenic activation of radionuclides.

\section{Dark Matter signature in RES-NOVA}
\label{sec:2}
The search for DM through direct detection experiments is a cornerstone of contemporary astroparticle physics~\cite{DM_review2024}. The RES-NOVA project possesses characteristics that may also make it a valuable tool in the search for DM, particularly for WIMPs and other dark matter candidates that may interact with ordinary baryonic matter.

The principle underlying DM detection in RES-NOVA is akin to that for neutrinos: the interaction of a DM particle with the nucleus of the target material, leading to a detectable recoil. Given the expected low interaction cross-section of DM, the choice of target material and the sensitivity of the detection apparatus are critical. The large neutron and atomic numbers of Pb make it an ideal target not only for studying CE$\nu$NS interactions but also for investigating DM candidates. Furthermore, the use of ultra-low background archaeological Pb by RES-NOVA significantly enhances the sensitivity for observing these types of weak interactions.

\begingroup\color{black}
We can give a preliminary estimate of the RES-NOVA sensitivity to SI DM interactions by requiring that the expected time-integrated event rate due to DM–nucleus scattering, $R_{DM}$, be equivalent to that from SN neutrinos via CE$\nu$NS, $R_{CE \nu NS}$. Under this assumption, we obtain:
\begin{eqnarray}
R_{CE \nu NS} &\sim& \sigma_{CE \nu NS} \cdot T({10~s}) \sim N^2 \cdot 10~s \\
R_{DM} &\sim& \sigma_{DM} \cdot T({1~y}) \sim A^2 \cdot 3\times 10^7~s \sim (2N)^2 \cdot 3\times 10^7~s\\
R_{DM} &\simeq& R_{CE \nu NS} \;\; \rightarrow \;\; \sigma_{DM} \sim 10^{-7} \cdot \sigma_{CE \nu NS}
\end{eqnarray}

where $T$ is the measurement time, $A$ and $N$ are the atomic and neutron number of the target nucleus, respectively. Representative values for $\sigma_{CE \nu NS}$ are shown in Fig.~\ref{fig:sigma}.

This simplified scaling argument suggests that, RES-NOVA while being sensitive to SN neutrinos, it can also probe SI DM cross-section to the 10$^{-45}$~cm$^2$ scale.
\endgroup

The expected signature of DM interactions in RES-NOVA would be nuclear recoils of a few keVs of energy, similar to those expected from neutrino interactions but with a potentially different temporal distribution and energy spectrum. The greatest difference lies in the nature of the investigated source: supernova-emitted neutrinos will result in a transient signal expected to completely fade away in around 10~s, whereas DM-induced nuclear recoils are a persistent signal. The annual modulation effect due to the motion of Earth relative to the Sun is believed to affect the DM interaction rate in the detector by around 5\%~\cite{Review_modulation}. 

The DM differential cross-section is described as follows~\cite{DM_review2024}:
\begin{equation}
\label{eq:dm_xsec}
\frac{d\sigma}{dE_R} = \frac{m_N}{2 v^2 \mu^2_{n}} \; \sigma_{\text{DM-N}} \cdot A^2 \cdot |F(q)|^2,
\end{equation}
where $\sigma_{\text{DM-N}}$ is the DM-nucleon cross-section, $v$ is the relative velocity between the DM particle and the detector, $\mu_n$ is the DM-nucleus reduced mass. The term $A^2$ represents the enhancement of the interaction due to coherence, where $A$ is the atomic number of the target material. Finally, $F(q)$ is the nuclear form factor, reflecting the loss of coherence in the interaction at a given momentum transfer $q$. This expression highlights the dependence of the detection probability on the properties of both the DM particle and the target nucleus.

RES-NOVA will use PbWO$_4$ crystals as the target material for both neutrino and DM detection~\cite{kg-scale}. In addition to the mentioned advantages of Pb, the inclusion of oxygen O in the crystal structure facilitates substantial momentum transfer from DM particles to the target. This results in higher energy nuclear recoils, thereby enhancing the experimental sensitivity to DM interactions over a broad mass range.
Given this experimental framework, RES-NOVA has the potential to probe a broad range of DM masses and interaction cross-sections that are complementary to existing searches. 

In addition to the SI DM interaction framework~\cite{Lewin:1995rx, Schumann:2019eaa}, RES-NOVA can also provide insights into the spin-dependent (SD) interaction channel through the use of isotopes like $^{207}$Pb (natural isotopic abundance of 22.1\%~\cite{IUPAP}, $J^{\pi}=1/2^-$~\cite{Kondev_2021}), but also $^{17}$O (natural isotopic abundance of 0.037\%~\cite{IUPAP}, $J^{\pi}=5/2^+$~\cite{Kondev_2021}). This type of interaction describes the DM interaction with unpaired nuclear spin~\cite{Engel:1992bf}, thus requiring a non-zero nuclear spin of target nuclei, as it is the case of the previously mentioned isotopes that have an odd mass number.
The SD interaction cross-section has similar formulation as the one for SI interactions (Eq.~\ref{eq:dm_xsec}), but with the complication that one needs to take into account the spin matrix element, dependent on the details of the nuclear structure~\cite{Engel:1992bf}:
\begin{equation}
\label{eq:dm_xsecSD}
\frac{d\sigma^{SD}}{dq^2} = \frac{\sigma^{SD}_{\text{DM-N}}}{3\mu_n v^2 }\frac{\pi}{(2J+1)}S_n(q),
\end{equation}
where $J$ is the target nuclear spin, and $S_n$ is the spin-matrix element for the interaction on neutrons. The computation of this last parameter can be quite complicated due to all possible multi-poles that complex heavy nuclei might have. In addition, the spin matrix elements is crucial for accurate calculations given its quenching beyond the limit of zero-momentum transfer~\cite{Bednyakov:2006ux}.
 
Within this framework, $^{207}$Pb is an outstanding target candidate: its low angular momentum prevents large quenching and its nuclear structure can be thought as a simple 2p$_{1/2}$ neutron hole within the otherwise complete shell structure of the doubly magic $^{208}$Pb, making it an ideal complement to the already well studied $^{73}$Ge and $^{29}$Si \cite{Kosmas:1997jm}.
Apart from these and other nuclides, listed in~\cite{Bednyakov:2006ux}, in all other cases a severe approximation of zero-momentum transfer is required.

These unique properties of $^{207}$Pb as a target for SD DM interactions complement those of $^{17}$O. Specifically, $^{17}$O is kinematically favorable for detecting light DM candidates, whereas $^{207}$Pb offers an accurate, momentum-dependent spin-structure description, making it well-suited for probing interactions involving heavier dark matter particles. In this work, we chose not to consider $^{183}$ W (natural isotopic abundance of 14.3\%~\cite{IUPAP}, $J^{\pi}=1/2^-$~\cite{Kondev_2021}), as it has the same nuclear spin and parity as $^{207}$Pb but a lower natural isotopic abundance, making it a less sensitive nuclear target.

\section{The RES-NOVA detector}\label{sec:det}
The RES-NOVA detector uses advanced thermal detection techniques, operating at cryogenic temperatures to enhance sensitivity and reduce noise. The core component is the absorber made of PbWO$_4$ with a very low heat capacity ($C$). The crystal absorber is thermally linked to a heat sink maintained at a constant mK temperature.

\begin{figure}[]
\centering
\includegraphics[width=.6\textwidth]{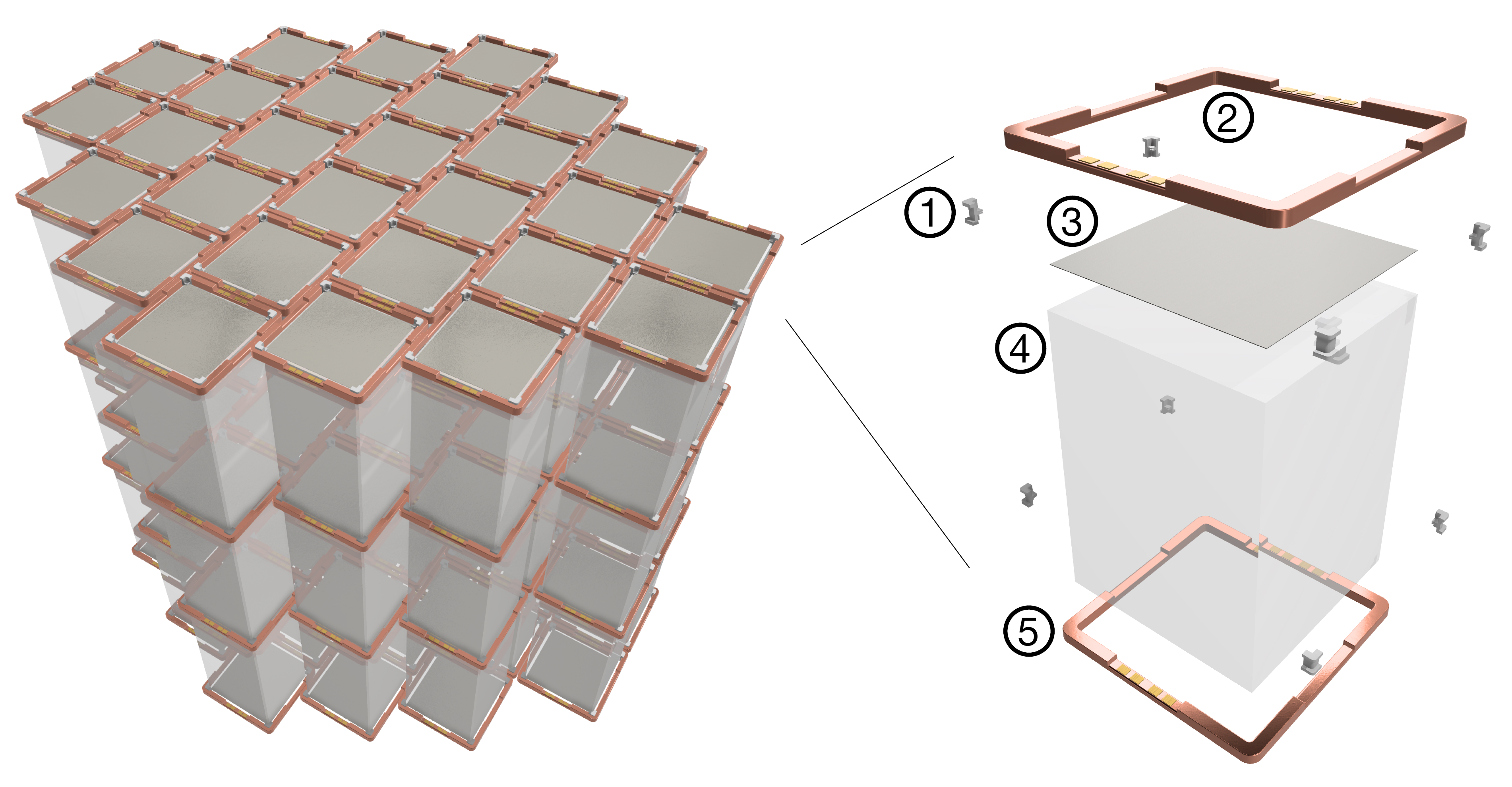}
\caption{Schematic view of the RES-NOVA detector demonstrator. It consists of 3 layers of 28 crystals each, for a total PbWO$_4$ mass of 170~kg. The details of the single detector module are shown on the right side of the figure: 1) PTFE holding support for the crystal and light detector, 2) Cu top frame of the structure, 3) scintillation light detector absorber, 4) PbWO$_4$ crystal, and 5) Cu bottom frame of the structure.\label{fig:det}}
\end{figure}

The experiment is planned to be installed at the underground Gran Sasso Laboratory (Italy), where it can benefit from the natural shielding of around 3800~m water equivalent from the Gran Sasso massif~\cite{G.Bellini_2012}. This is the ideal site for operating detectors in low-noise conditions.

When a particle (e.g. neutrino or DM) interacts with the absorber, it deposits energy ($\Delta E$), leading to a detectable temperature increase in the absorber, $\Delta T$. This temperature change is connected to the heat capacity of the absorber: $\Delta T \sim \Delta E/C$.
At cryogenic temperatures, the heat capacity $C$ is significantly reduced, closely following a Debye model of lattice dynamics, where $C(T)$ varies as $T^3$. This reduction in heat capacity is crucial for detecting minimal energy depositions, which translate into measurable temperature changes. In this context, the design of suitable temperature sensors is vital for achieving the required sensitivity and detecting the low-energy deposit induced by CE$\nu$NS and DM particle interactions.

RES-NOVA has successfully adopted Transistion Edge Sensors (TESs) as thermal sensors for the operation of eV-scale proof-of-principle cryogenic detectors made of PbWO$_4$ \cite{FerreiroIachellini:2021qgu}. This detector was characterized by a small absorber mass, about 15~g. The final RES-NOVA detector demonstrator is planning to achieve an energy threshold of about 1~keV, while operating kg-scale absorbers. \textcolor{black}{Following the prescriptions of~\cite{Franz} and the experimental scaling law presented in~\cite{Strauss:2017cam}, we expect to achieve the energy threshold target. In this regard, an extensive R\&D activity is currently carried out by the collaboration to show the feasibility of this goal.}
The detector demonstrator will have a total mass of about 170~kg, corresponding to 84~PbWO$_4$ units. This detector demonstrator will pave the way for RES-NOVA-1, which will have a mass of 1.8~t and its future extensions~\cite{Pattavina:2020cqc}.

The detector single unit, named module, is composed by a PbWO$_4$ crystal and a light detector (LD), namely a thin Ge absorber facing the crystal. This absorber serves as cryogenic calorimeter for the detection of the scintillation light produced by the main absorber~\cite{Beeman2013}. The simultaneous read-out of heat and light allows for a particle identification on a event-by-event base~\cite{Beeman:2012wz}. Each of these absorbers is held in place by a set of PTFE holders that also act as weak thermal link to the heat bath.

A preliminary detector's design is shown in Fig.~\ref{fig:det}. This is composed by 84 modules arranged in a tightly packed configuration of three layers, each hosting 28 modules. We consider this design as preliminary given that the final crystal's dimensions are not yet defined, as a specific R\&D on the crystal growth procedure is ongoing in order to increase the cross-sectional area of the crystals, from 17~cm$^2$ to 25~cm$^2$. Despite this ongoing activity, the whole detector active volume is defined to be around (30~cm)$^3$. The parameter that may change is the number of modules, however preliminary Monte Carlo studies, addressing this aspect, have shown that there is no significant impact in the overall background expectations if the total number of detectors is changed. Details on the expected background level in Region of Interest (RoI) are described in the following section.

\section{Background predictions}\label{sec:bkg}
\begin{figure}[]
\centering
\includegraphics[width=.6\textwidth]{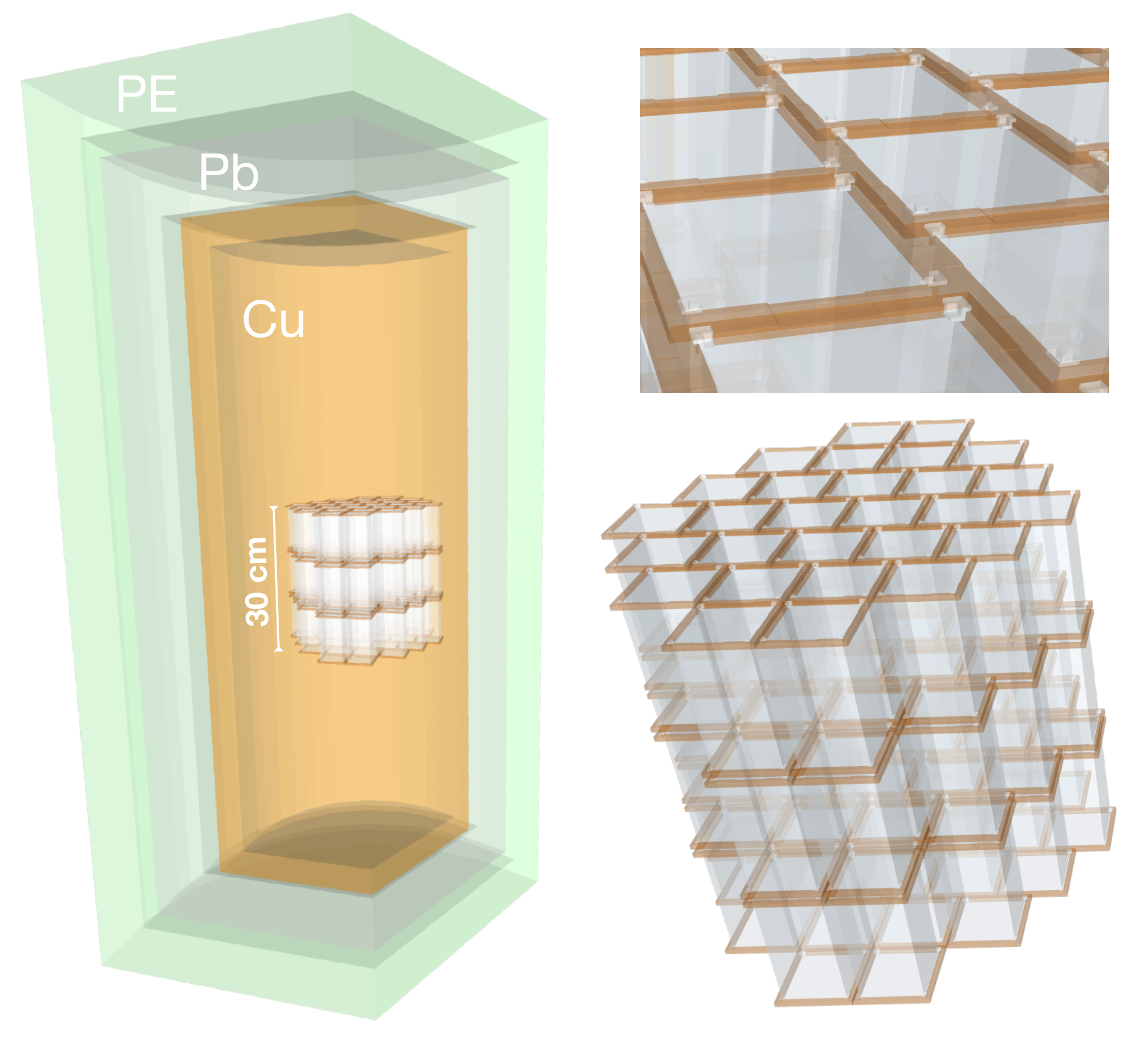}
\caption{Rendering of the RES-NOVA experimental set-up, as implemented in the Monte Carlo simulations. The detector is protected against environmental background sources thorough a set of shields made of: polyethylene (PE), lead (Pb) and copper (Cu). The detector is made of 84 PbWO$_4$ crystals arranged in three layers, and hold into position by a Cu structure and PTFE clamps. More details are provided in Sec.~\ref{sec:det}. \label{fig:MCdet}}
\end{figure}

A preliminary background projection for the experiment was outlined for the first time in \cite{Pattavina_2021}. In that work, the expected background level in the RoI and the physics potential of RES-NOVA phase 1, was presented. This detector has a total volume of (60~cm)$^3$ for a total of about 500~modules.
Starting from a similar design, but with reduced active volume (see Fig.~\ref{fig:MCdet}), we have extrapolated the background level achievable with the RES-NOVA demonstrator detector. For the sake of consistency and caution reasoning, we have not only included the same sources of background assumed in the previously mentioned work, but we have extended the list of background sources considered. The ultimate goal is to achieve a more precise understanding of all potential background sources in the experiment that could limit its sensitivity to detecting particles from the cosmos, whether they are neutrinos or DM candidates.

The background sources considered in this work are categorized according to their location: external (\textit{External}), in the detector shielding (\textit{Shields}), inside the \textit{Detector} (PbWO$_4$ + Structure), and on the \textit{Surface} of Cu/PTFE components facing the crystals, and on the crystals themselves. In the first class, \textit{External}, we accounted for contributions from cosmic rays, namely muons, and environmental neutrons and gammas in the underground laboratory. The \textit{Shields} class includes background sources coming from detector shielding materials. In this case, we have assumed a shielding of (from the outside to the detector inside): 15~cm of polyethylene (PE), 15~cm of \textcolor{black}{modern} Pb, and 7~cm of high-purity Cu. This last layer includes also the thermal radiation shielding directly facing the detector (Cu tiles). The \textit{Detector} class includes bulk contaminations from the crystals, the Cu structure and the PTFE holders. Here, we neglect the products of cosmogenic activation, which were evaluated as a secondary contributor~\cite{kg-scale}. Finally, in the \textit{Surface} class, we include contaminations in different components of the detector and infrastructure. We modeled these as a double layer of contaminants with an exponential profile with two characteristics depths, one superficial at 10~nm, and one sub-superficial at 10~$\mu$m~\cite{Clemenza:2011zz}.

We account for naturally occurring radioactive contaminations produced by the decay chains of $^{232}$Th and $^{238}$U, as well as the sub-chain of $^{210}$Pb. Secondary products of the interactions of these nuclides with the detector / infrastructure components (e.g. electrons, Bremsstrahlung radiation) are also considered.
In Tab.~\ref{tab:activities}, we include a summary of the different background sources considered for the estimation of the background. All the values reported as limits (90\% C.L.) were considered as values for the Monte Carlo simulations.

\begin{table*}[t]
    \centering
    \begin{tabular}{|l|l|l|l|l|}
        \hline
        \textbf{Location} & \textbf{Component} & \textbf{Source} & \textbf{Activity [mBq/kg] ([mBq/cm$^2$])} & \textbf{Refs.} \\ \hline
        \multirow{8}{*}{\textbf{Detector}} & \multirow{4}{*}{\textbf{PbWO$_4$}} & $^{232}$Th & $< 2.3 \times 10^{-1}$ & \cite{Belli:2020qqc} \\
                                           &                            & $^{238}$U & $< 7.0 \times 10^{-2}$ & \cite{Belli:2020qqc}  \\ 
                                           &                            & $^{210}$Pb & $< 7.1 \times 10^{-1}$ & \cite{Pattavina:2019pxw} \\
                                           &                            & $^{40}$K & $< 9.0 \times 10^{-2}$ & \cite{Belli:2020qqc} \\ \cline{2-5}
                                           & \multirow{2}{*}{\textbf{Cu structure}}      & $^{232}$Th & $< 4.1 \times 10^{-5}$ & \cite{Data-driven} \\ 
                                           &                            & $^{238}$U & $< 1.7 \times 10^{-5}$ & \cite{Data-driven} \\ \cline{2-5}
                                           & \multirow{2}{*}{\textbf{PTFE holders}} & $^{232}$Th & $< 6.1 \times 10^{-3}$ & \cite{Data-driven} \\ 
                                           &                            & $^{238}$U & $< 2.2 \times 10^{-2}$ & \cite{Data-driven} \\ \hline
        \multirow{9}{*}{\textbf{Shields}} & \multirow{2}{*}{\textbf{Polyethylene}} & $^{232}$Th & $< 9.4 \times 10^{-2}$ & \cite{Aprile:2010zz} \\ 
                                          &                                        & $^{238}$U & $< 0.23$ & \cite{Aprile:2010zz} \\ \cline{2-5}
                                          & \multirow{3}{*}{\textbf{Pb shield}} & $^{232}$Th & $< 2.3$ & \cite{Data-driven} \\ 
                                          &                                        & $^{238}$U & $< 3.5$ & \cite{Data-driven} \\ 
                                          &                                        & $^{210}$Pb & $ 3 \times 10^{5}$ & \cite{Data-driven} \\ \cline{2-5}
                                          & \multirow{2}{*}{\textbf{Cu shield}}   & $^{232}$Th & $< 4.1 \times 10^{-2}$ & \cite{Data-driven} \\ 
                                          &                                        & $^{238}$U & $< 1.7 \times 10^{-2}$ & \cite{Data-driven} \\ \hline
        \multirow{10}{*}{\textbf{Surface}} & \multirow{2}{*}{\textbf{PbWO$_4$}} & $^{210}$Pb - 10~$\mu$m & $7.4 \times 10^{-6}$ & \cite{Data-driven} \\ 
                                           &                            & $^{210}$Pb - 0.01~$\mu$m & $ 8.7 \times 10^{-5}$ & \cite{Data-driven} \\ \cline{2-5}
                                           & \multirow{5}{*}{\textbf{Cu structure/Tiles}} & $^{232}$Th - 10~$\mu$m & $(1.2 \pm 0.1) \times 10^{-5}$ & \cite{Data-driven} \\ 
                                           & \multirow{5}{*}{\textbf{PTFE holders}} & $^{238}$U - 10~$\mu$m & $(8.3 \pm 0.7) \times 10^{-6}$ & \cite{Data-driven} \\ 
                                           &                            & $^{210}$Pb - 10~$\mu$m & $ 1.2 \times 10^{-4}$ & \cite{Data-driven} \\ 
                                           &                            &  $^{232}$Th - 0.01~$\mu$m & $(1.4 \pm 0.1) \times 10^{-6}$ & \cite{Data-driven} \\ 
                                           &                            &  $^{238}$U - 0.01~$\mu$m & $(1.2 \pm 0.1) \times 10^{-6}$ & \cite{Data-driven} \\ 
                                           &                            & $^{210}$Pb - 0.01~$\mu$m & $(4.1 \pm 0.1) \times 10^{-4}$ & \cite{Data-driven} \\\cline{2-5}
                                           & \multirow{1}{*}{\textbf{Cu shield}} & $^{210}$Pb - 0.01~$\mu$m & $1 \times 10^{-2}$ & \cite{Data-driven} \\ \hline \hline
       \multirow{3}{*}{\textbf{External}} &  & neutrons & $3.9 \times 10^{-6} \, \text{cm}^{-2}\text{s}^{-1}$ & \cite{Wulandari:2003cr} \\
         &  & muons & $3.1 \times 10^{-8} \, \text{cm}^{-2}\text{s}^{-1}$ & \cite{Data-driven} \\
          &  & gammas & $4.0 \times 10^{-1} \, \text{cm}^{-2}\text{s}^{-1}$ & \cite{LNGS-gamma} \\ \hline
    \end{tabular}
    \caption{List of background sources considered for the RES-NOVA background prediction. The table includes the location, specific activity (measured limits at 90\% C.L. and values) for bulk and surface contaminations, as well as fluxes for environmental sources.}
    \label{tab:activities}
\end{table*}

All the results of the simulations presented here are run with the Monte Carlo code (\textit{Arby}) based on the \texttt{GEANT4} toolkit~\cite{Agostinelli:2002hh}. 

\begin{figure*}[h]
  \includegraphics[width=.49\textwidth]    {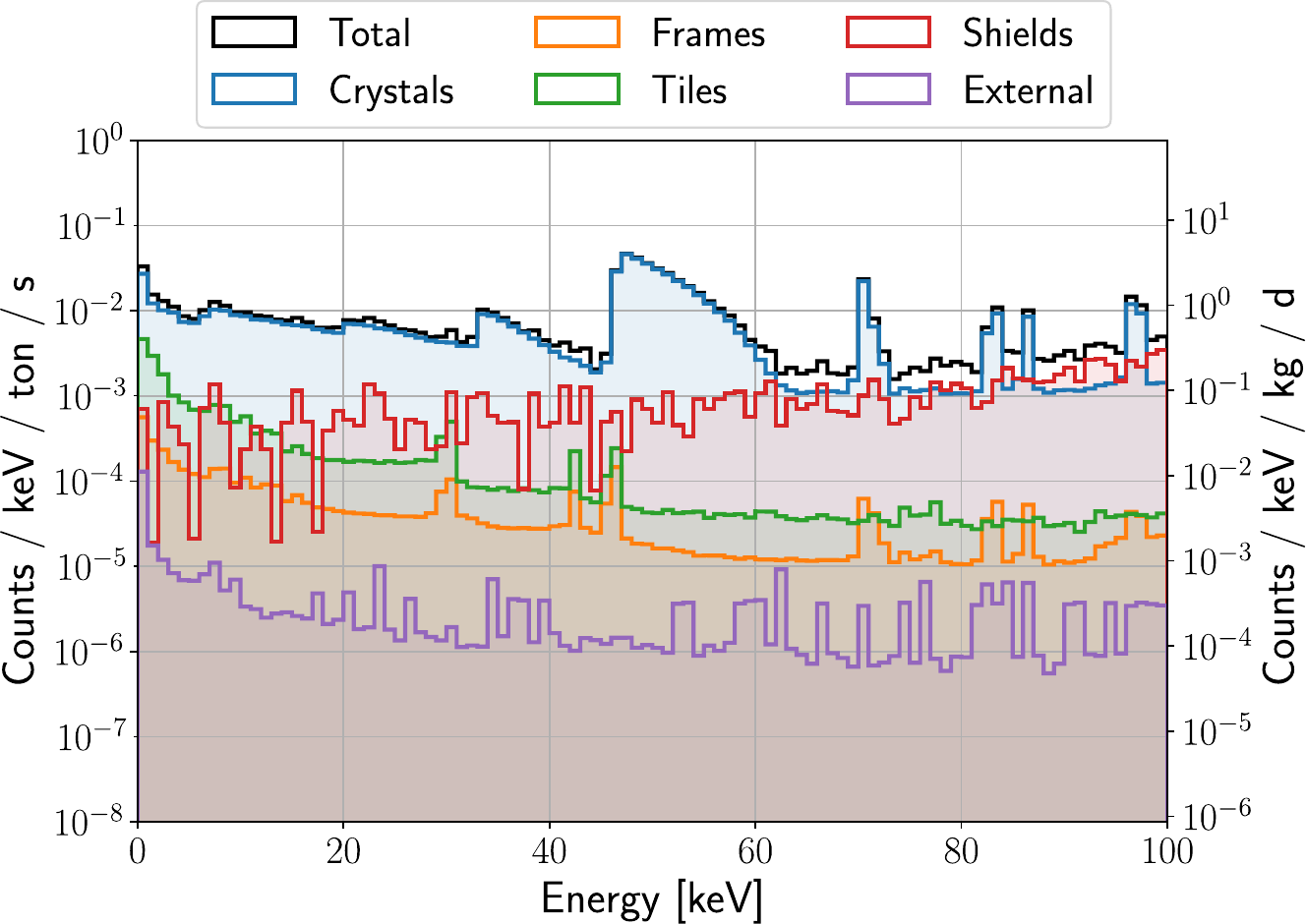} \;\hfill
  \includegraphics[width=.49\textwidth]    {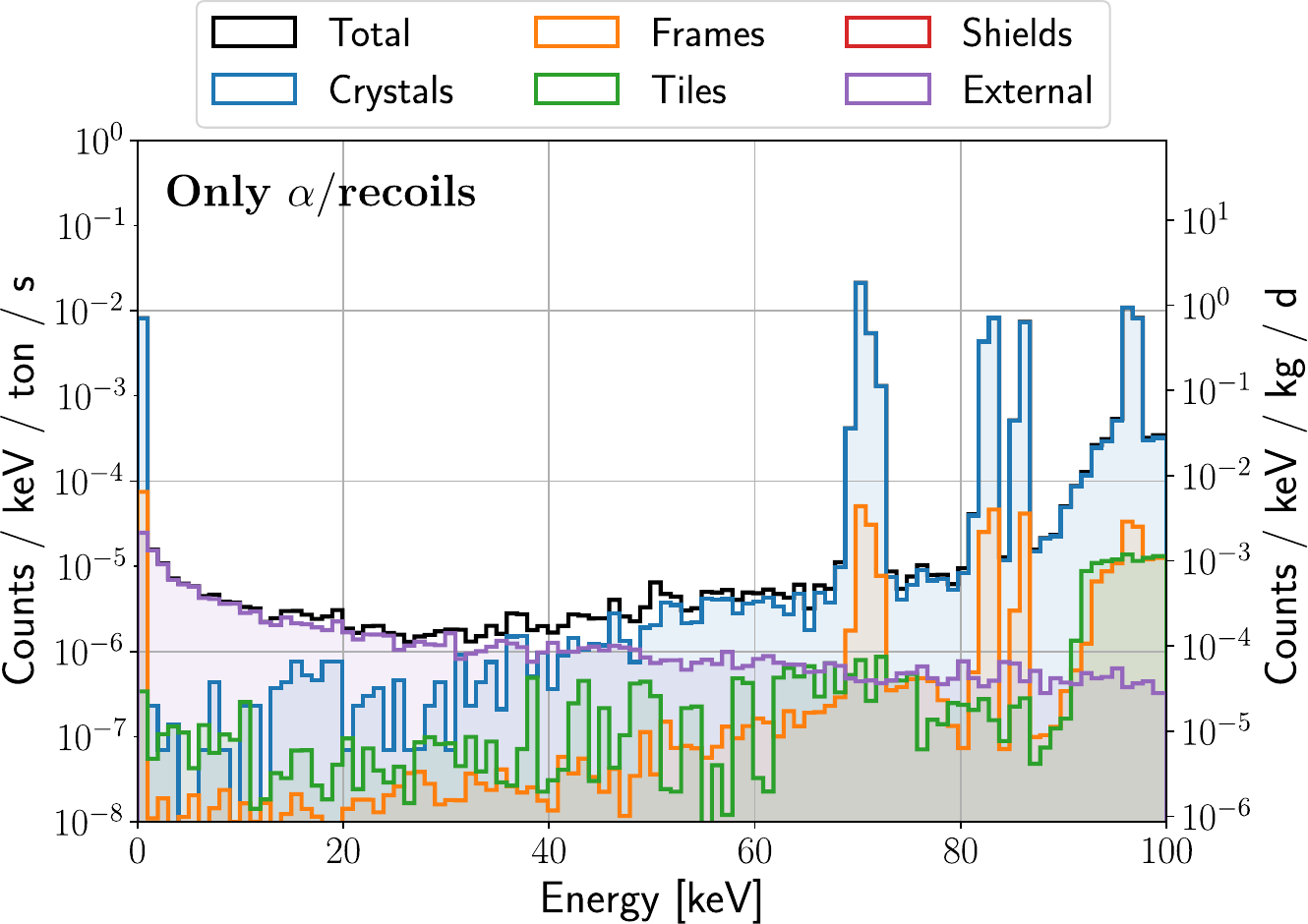}
  \caption{Background prediction in the region of interest when the RES-NOVA detector is operated in anti-coincidence mode. RES-NOVA has a total active volume of about (30~cm)$^3$, and a mass of about 170~kg of PbWO$_4$ crystals produced from archaeological Pb. In the left (right) plot, the total energy spectrum produced by background sources in the experimental set-up is shown. In the right plot, the total energy spectrum (same as in the left plot) where a 100\% rejection of $e^-$ and $\gamma$ events is performed, leaving only nuclear recoils and alphas.\label{fig:spectra}}
\end{figure*}

The prediction of the total background for the RES-NOVA demonstrator is presented in Fig.~\ref{fig:spectra}, where we show the detector response to the different classes of background sources. As detector operating parameters, we considered: detector energy resolution of 200~eV, time resolution of 0.1~ms, trigger and data acquisition window of 500~ms and detector counting rate 0.1~mHz. These specifications are required to provide realistic detector response and data processing. In fact, we selected only events in anti-coincidence among the different detectors, where the events are required to be spaced at least 500~ms (an acquisition window).

In Fig.~~\ref{fig:spectra}, we assumed two different operating configurations: one where no particle identification or discrimination is performed (PbWO$_4$ read-out only), and the other where a 100\% efficient rejection of $e^-$ and $\gamma$ events is performed (PbWO$_4$ and LD read-out). We adopt this approach to show the two extreme cases of conservative and very optimistic background predictions. In the first case, we assume that there is no improvement in the intrinsic crystal contamination levels, namely the dominant background, despite the fact that those are limits and not measured values. In the second, we assume a highly efficient light collection as well as a high light yield, to enable a full rejection of $e^-$ and $\gamma$ events. In the best case scenario, we cannot improve the background of the experiment in the RoI by more than a factor of 10$^3$ when we also perform the read-out of the PbWO$_4$ scintillation light.

The results of the simulations point to the crystal bulk contaminations as the main component of the background. In fact, the collaboration is currently addressing many efforts in benchmarking the production line of the PbWO$_4$ crystals.
The baseline of the experiment is to achieve a background of 10$^{-3}$~c/keV/ton/s \footnote{Equivalent to 0.086~c/keV/kg/d.} when the detector is operated in coincidence mode, e.g. more than one module triggers an event in a time window of about 1~s, for more details about the definition of coincidence mode in RES-NOVA see Ref.~\cite{Pattavina_2021}. 

\section{Methodology}
The experimental sensitivity to a DM candidate is conventionally expressed as the smallest DM-nucleon cross-section that in 9 out of 10 realizations of the experiment induces a number of events that cannot be ascribed to Poissonian (upper) fluctuations of the background at 90\% confidence level. In formal terms, that means finding the smallest DM-nucleon interaction cross-section $\sigma_0$ such that for any given experimental outcome $X$:
\newcommand{\probP}{\text{I\kern-0.15em P}}
\begin{equation}
\centering
\label{eq:cls}
\frac{\probP(\sigma(X) < \sigma_0 \;|\; \sigma_{\text{DM-N}} > 0)}{\probP(\sigma(X) < \sigma_0 \;|\; \sigma_{\text{DM-N}} = 0)} \leq \alpha \;\;\text{where}\;\; 1-\alpha=0.9,
\end{equation}
where $\probP(\sigma(X) <\sigma_0 \;|\; \sigma_{\text{DM-N}}> 0)$ is the probability of observing a non-vanishing interaction cross-section under the assumption of a truly non-vanishing interaction cross-section, and $\probP(\sigma(X) < \sigma_0 \;|\; \sigma_{\text{DM-N}} = 0)$ is the probability of observing a non-vanishing interaction cross-section under the background-only hypothesis.
In order to determine these probabilities, we simulate the expected event rate of a background-only realization of RES-NOVA. 

For the \textit{i}-th bin in the energy spectrum we compute the number of expected events $n_i$ as the sum of the background counts $b_i$ and the counts due to DM interactions $s_i$ multiplied by a strength parameter $\mu$ (the latter is independent from the index \textit{i}):
\begin{equation}
\centering
\label{eq:dm_excl}
n_i = \mu s_i+b_i.
\end{equation}
In Eq.~\ref{eq:dm_excl}, we can then identify the case in which the signal is absent (null-hypothesis, $\mu=0$) and a potential discovery ($\mu=1$). We use the upper-limit q$_\mu$ statistics~\cite{Cowan:2010js} to set a limit (and relative uncertainty) on $\mu$ that, being proportional to the number of DM interactions, translates to the limit on the cross-section $\sigma_0$. 

For the calculation of the upper limit on $\mu$ we used the package \texttt{PyHF}~\cite{pyhf, pyhf_joss}. In order to compute the expected recoil rate $s_i$ in our targets, we have adapted the Xenon Collaboration \texttt{wimprates} package~\cite{jelle_aalbers_2023_7636982} to include Pb, W and O isotopes (for the SD calculation, we have included the spin nuclear structure of $^{207}$Pb discussed in Sec.~\ref{sec:2}). The halo model parameters are taken from~\cite{Baxter:2021pqo}: local DM density $\rho_{DM}=0.3$~GeV/c$^2$/cm$^3$, galactic escape velocity $v_{esc}=544$~km/s, local standard rest velocity $v_{lab}=238$~km/s, solar peculiar velocity $v_{\odot}=(11.1, 12.2, 7.3)$~km/s, average galactocentric Earth speed $v_\oplus=29.8$~km/s. The background signal $b_i$ is the one presented in Sec.~\ref{sec:bkg}.

\section{Sensitivity projections}\label{sec:projection}
We have explored the potential of RES-NOVA to DM particles in the local halo. Our simulations assume a total detector exposure of 170~kg$\cdot$y and a nuclear recoil energy resolution of 200~eV, corresponding to a 50\% trigger efficiency at a nuclear recoil energy of 1~keV (with a hard cut at 200~eV)— the target energy threshold of the RES-NOVA experiment~\cite{Pattavina:2020cqc}. As input energy spectrum for our sensitivity studies, we simply scaled the results presented in Fig.~\ref{fig:spectra} for the detector exposure.
\begin{figure}[]
\centering
  \includegraphics[width=.6\textwidth]    {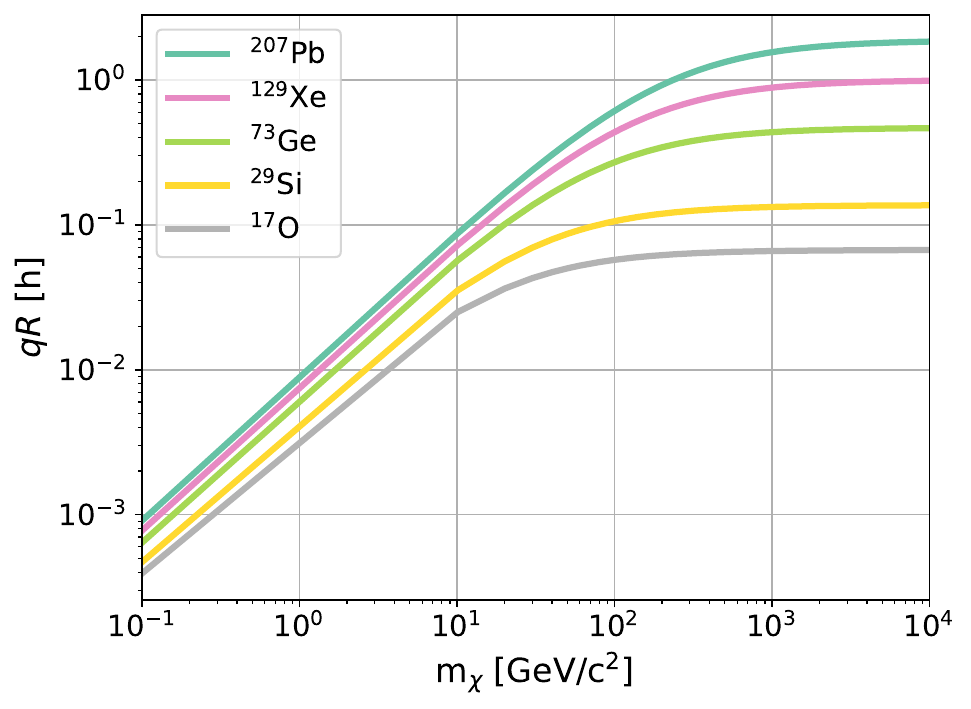}
  \caption{$qR$ in units of $h$ ($q$ is the transferred momentum and $R$ the nuclear radius) for the more common isotopes used as target for SD Dark Matter particle interactions. The $^{17}$O and $^{207}$Pb targets adopted in this work lie at the two extremes of the plot, where the zero-momentum approximation holds at its best and at its worst.}\label{fig:qR}
\end{figure}

Sensitivity calculations were performed within the standard thermal relic WIMP framework~\cite{Baxter:2021pqo}, accounting for spin-independent and spin-dependent interactions with the target. For this analysis, we considered the two mentioned scenarios: a conservative scenario, where no particle identification is implemented, and an optimistic one, where RES-NOVA achieves 100\% efficiency in rejecting non-nuclear recoil interactions.

It is worth emphasizing that this is the first time projections are presented for Pb as active target material. In particular, for the spin-dependent case, the known nuclear spin structure of $^{207}$Pb provides more robust results at non-negligible momentum transfer compared to other compounds~\cite{CRESST:2019jnq, CRESST:2024cpr}, where spin-dependent sensitivity is exclusively dependent on $^{17}$O and is therefore limited to the zero-momentum transfer approximation. This approximation holds when the quantity $qR\ll1$, where $q$ is the transferred momentum and $R$ the nuclear radius. Some care is needed when making such assumption, as it may not always be the case that the transferred momenta are small in comparison to the nuclear size~\cite{Ressell:1997kx}. In order to quantify the goodness of this approximation, we plot the quantity $qR$ determined with the maximum possible $q$ (i.e. a frontal hit of a DM particle traveling at the escape velocity) as a function of the DM particle mass (Fig.~\ref{fig:qR}) and we observe that the larger the target nucleus, the more important the accounting for its nuclear spin structure.

\begin{figure*}[]
  \includegraphics[width=.49\textwidth]    {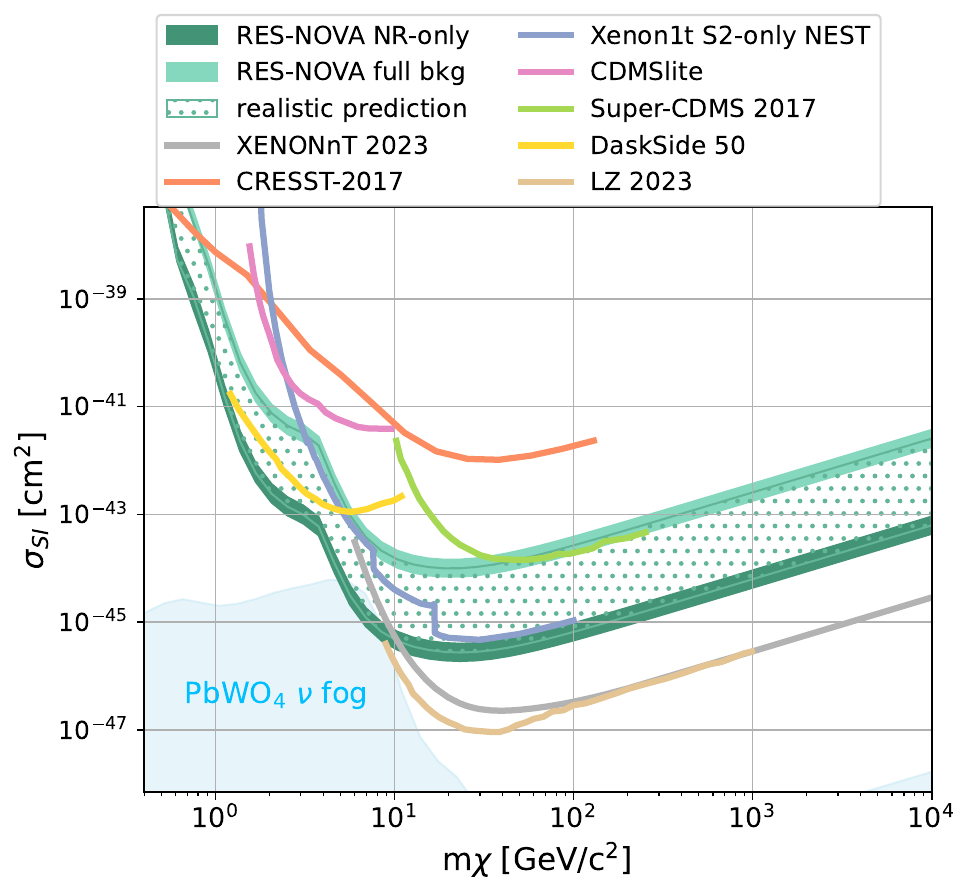} \hfill
  \includegraphics[width=.49\textwidth]    {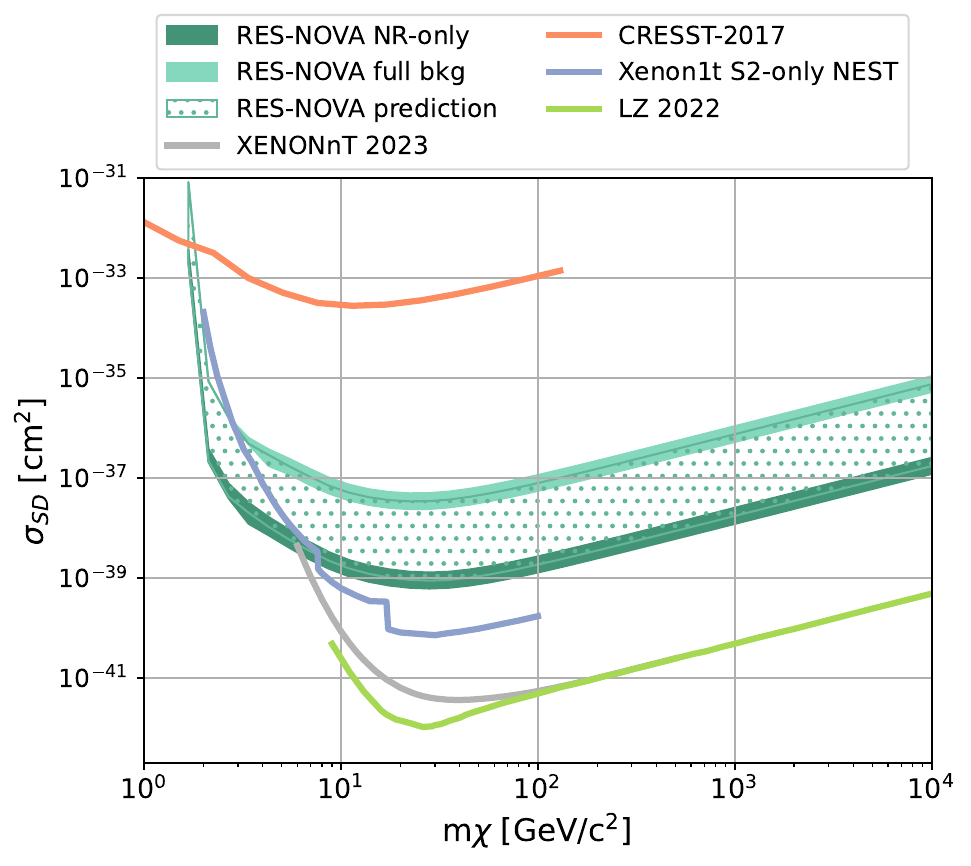}
  \caption{RES-NOVA sensitivity predictions on the cross-section for dark matter particles scattering off nucleons (spin-independent, \textit{left}) and scattering off neutrons (spin-dependent, \textit{right}) after 1 year of data taking, equivalent to an exposure of 170~kg$\cdot$y.  The RES-NOVA predictions consider the conservative (lighe green band) and optimistic (dark green band) scenarios of the background level in the region of interest. Exclusion limits from other searches are taken from Refs.~\cite{CRESST:2019jnq, DarkSide:2018bpj, XENON:2019gfn, LZ:2022lsv, SuperCDMS:2017mbc, PhysRevLett.112.041302, XENON:2023cxc}. The neutrino fog was determined according to Ref.~\cite{Billard:2013qya}.\label{fig:limit}}
\end{figure*}

Our projections for SI and SD DM interactions for a continuous operation of RES-NOVA over 1~y are shown in Fig.~\ref{fig:limit}. In the same figure we also show the expected neutrino background produced by solar and atmospheric neutrino interactions in PbWO$_4$ crystals. The so-called \textit{neutrino fog}~\cite{Fog} is calculated as the maximum achievable sensitivity, independently of the energy threshold (we tested energy thresholds between $10^{-3}$~eV and 10~keV), without detecting any neutrino (at 90\% CL) from the Sun (below 1~GeV/$c^2$ of mass) or from the atmosphere (above 1~GeV/$c^2$ of mass).
As expected, the background rejection capability shows a significant impact on the achievable DM sensitivity. If we take for example a WIMP of 30~GeV/c$^2$ of mass, interacting according to the SI model, we expect an improvement of more than 1 order of magnitude with an exposure of 2 ton$\cdot$year, both in the optimistic and in the pessimistic scenario. At that point, the sensitivity will be background limited and only moderate improvement can be achieved by increasing the exposure. 
The sensitivity predictions presented in Fig.~\ref{fig:limit} show the potential of a low-background cryogenic detector made from archaeological Pb. The synergistic combination of heavy and light targets (Pb and O respectively) allows for the exploration of a broad range of DM masses that covers 4-orders of magnitude for SI interactions. RES-NOVA in this scenario can provide complementary results to those already achieved by Xenon-based detectors, a leading technology in this sector. Similar results, but with limited impact, can also be achieved during the study of SD DM interactions on neutrons.

\section{Summary}
This work presents the DM detection potential of the RES-NOVA experiment, effectively extending its physics reach beyond SN neutrino detection.

RES-NOVA is the first experiment employing Pb as an active material in a low-background setup. The exceptional properties of archaeological Pb, namely its radiopurity and high nuclear mass, enable RES-NOVA to achieve DM detection sensitivities comparable and complementary to those of ongoing direct detection experiments~\cite{LZ:2022lsv,XENON:2023cxc,PandaX_2021, Agnese:2017jvy,CRESST:2019jnq}. After 1~y of data taking in the best background configuration, RES-NOVA can probe DM interaction cross-section with baryonic matter down to 1$\times 10^{-43}$~cm$^2$ for DM masses below the Lee-Weinberg bound of 2~GeV/c$^2$~\cite{Lee-Weinberg}. The most stringent limit on the cross-section is 2$\times 10^{-46}$~cm$^2$ achieved for candidates with 20~GeV/c$^2$ of mass. Furthermore, the relatively high abundance of $^{207}$Pb provides an interesting opportunity to explore SD DM interactions, in a sector that has been studied less extensively.

This work also outlines the development of an improved Monte Carlo background model for the RES-NOVA experimental configuration. The impact of this background model on the search for DM particles scattering off nuclei is carefully quantified. Furthermore, the potential to improve background rejection through the simultaneous reading of heat and light in PbWO$_4$ is explored when used as a cryogenic detector. Our simulations demonstrate that the RES-NOVA demonstrator, operating over a 1~y period, achieves a sensitivity that remains not limited by the background level. Notably, scaling up the exposure by an order of magnitude could yield even more promising results without reaching background limitations, suggesting significant potential for future expansions of the experimental setup. These findings establish a strong foundation for the development of larger-scale RES-NOVA detectors, which could substantially enhance the exploration of the dark matter parameter space while maintaining the advantages of the current design.

\acknowledgments
This work received funding from the EU Horizon Europe program through grant ERC-101087295-RES-NOVA.
This work makes use of the \texttt{Arby} software for \texttt{Geant4}-based Monte Carlo simulations, which was developed within the framework of the Milano-Bicocca activities and is maintained by O.~Cremonesi and S.~Pozzi.
We are grateful to the University of Milano-Bicocca (UNIMIB) and the Istituto Nazionale di Fisica Nucleare (INFN) for actively supporting the collaboration. V.V.K., D.V.K. and O.G.P. were supported in part by the National Research Foundation of Ukraine Grant No. 2023.03/0213.

\bibliography{apssamp}

\begin{thebibliography}{65}%
\makeatletter
\providecommand \@ifxundefined [1]{%
 \@ifx{#1\undefined}
}%
\providecommand \@ifnum [1]{%
 \ifnum #1\expandafter \@firstoftwo
 \else \expandafter \@secondoftwo
 \fi
}%
\providecommand \@ifx [1]{%
 \ifx #1\expandafter \@firstoftwo
 \else \expandafter \@secondoftwo
 \fi
}%
\providecommand \natexlab [1]{#1}%
\providecommand \enquote  [1]{``#1''}%
\providecommand \bibnamefont  [1]{#1}%
\providecommand \bibfnamefont [1]{#1}%
\providecommand \citenamefont [1]{#1}%
\providecommand \href@noop [0]{\@secondoftwo}%
\providecommand \href [0]{\begingroup \@sanitize@url \@href}%
\providecommand \@href[1]{\@@startlink{#1}\@@href}%
\providecommand \@@href[1]{\endgroup#1\@@endlink}%
\providecommand \@sanitize@url [0]{\catcode `\\12\catcode `\$12\catcode `\&12\catcode `\#12\catcode `\^12\catcode `\_12\catcode `\%12\relax}%
\providecommand \@@startlink[1]{}%
\providecommand \@@endlink[0]{}%
\providecommand \url  [0]{\begingroup\@sanitize@url \@url }%
\providecommand \@url [1]{\endgroup\@href {#1}{\urlprefix }}%
\providecommand \urlprefix  [0]{URL }%
\providecommand \Eprint [0]{\href }%
\providecommand \doibase [0]{https://doi.org/}%
\providecommand \selectlanguage [0]{\@gobble}%
\providecommand \bibinfo  [0]{\@secondoftwo}%
\providecommand \bibfield  [0]{\@secondoftwo}%
\providecommand \translation [1]{[#1]}%
\providecommand \BibitemOpen [0]{}%
\providecommand \bibitemStop [0]{}%
\providecommand \bibitemNoStop [0]{.\EOS\space}%
\providecommand \EOS [0]{\spacefactor3000\relax}%
\providecommand \BibitemShut  [1]{\csname bibitem#1\endcsname}%
\let\auto@bib@innerbib\@empty
\bibitem [{\citenamefont {Billard}\ \emph {et~al.}(2022)\citenamefont {Billard} \emph {et~al.}}]{Billard_2022}%
  \BibitemOpen
  \bibfield  {author} {\bibinfo {author} {\bibfnamefont {J.}~\bibnamefont {Billard}} \emph {et~al.},\ }\bibfield  {title} {\bibinfo {title} {Direct detection of dark matter {APPEC} committee report},\ }\href {https://doi.org/10.1088/1361-6633/ac5754} {\bibfield  {journal} {\bibinfo  {journal} {Rep. Prog. Phys.}\ }\textbf {\bibinfo {volume} {85}},\ \bibinfo {pages} {056201} (\bibinfo {year} {2022})}\BibitemShut {NoStop}%
\bibitem [{\citenamefont {Navas}\ \emph {et~al.}(2024)\citenamefont {Navas} \emph {et~al.}}]{ParticleDataGroup}%
  \BibitemOpen
  \bibfield  {author} {\bibinfo {author} {\bibfnamefont {S.}~\bibnamefont {Navas}} \emph {et~al.} (\bibinfo {collaboration} {{Particle Data Group}}),\ }\bibfield  {title} {\bibinfo {title} {Review of particle physics},\ }\href {https://doi.org/10.1103/PhysRevD.110.030001} {\bibfield  {journal} {\bibinfo  {journal} {Phys. Rev. D}\ }\textbf {\bibinfo {volume} {110}},\ \bibinfo {pages} {030001} (\bibinfo {year} {2024})}\BibitemShut {NoStop}%
\bibitem [{\citenamefont {Bertone}\ \emph {et~al.}(2005)\citenamefont {Bertone}, \citenamefont {Hooper},\ and\ \citenamefont {Silk}}]{Cirelli:2024ssz}%
  \BibitemOpen
  \bibfield  {author} {\bibinfo {author} {\bibfnamefont {G.}~\bibnamefont {Bertone}}, \bibinfo {author} {\bibfnamefont {D.}~\bibnamefont {Hooper}},\ and\ \bibinfo {author} {\bibfnamefont {J.}~\bibnamefont {Silk}},\ }\bibfield  {title} {\bibinfo {title} {{Particle dark matter: Evidence, candidates and constraints}},\ }\href {https://doi.org/10.1016/j.physrep.2004.08.031} {\bibfield  {journal} {\bibinfo  {journal} {Phys. Rept.}\ }\textbf {\bibinfo {volume} {405}},\ \bibinfo {pages} {279} (\bibinfo {year} {2005})},\ \Eprint {https://arxiv.org/abs/hep-ph/0404175} {arXiv:hep-ph/0404175} \BibitemShut {NoStop}%
\bibitem [{\citenamefont {Pattavina}\ \emph {et~al.}(2020)\citenamefont {Pattavina}, \citenamefont {Ferreiro~Iachellini},\ and\ \citenamefont {Tamborra}}]{Pattavina:2020cqc}%
  \BibitemOpen
  \bibfield  {author} {\bibinfo {author} {\bibfnamefont {L.}~\bibnamefont {Pattavina}}, \bibinfo {author} {\bibfnamefont {N.}~\bibnamefont {Ferreiro~Iachellini}},\ and\ \bibinfo {author} {\bibfnamefont {I.}~\bibnamefont {Tamborra}},\ }\bibfield  {title} {\bibinfo {title} {{Neutrino observatory based on archaeological lead}},\ }\href {https://doi.org/10.1103/PhysRevD.102.063001} {\bibfield  {journal} {\bibinfo  {journal} {Phys. Rev. D}\ }\textbf {\bibinfo {volume} {102}},\ \bibinfo {pages} {063001} (\bibinfo {year} {2020})},\ \Eprint {https://arxiv.org/abs/2004.06936} {arXiv:2004.06936 [astro-ph.HE]} \BibitemShut {NoStop}%
\bibitem [{\citenamefont {Pirro}\ and\ \citenamefont {Mauskopf}(2017)}]{Pirro:2017ecr}%
  \BibitemOpen
  \bibfield  {author} {\bibinfo {author} {\bibfnamefont {S.}~\bibnamefont {Pirro}}\ and\ \bibinfo {author} {\bibfnamefont {P.}~\bibnamefont {Mauskopf}},\ }\bibfield  {title} {\bibinfo {title} {{Advances in Bolometer Technology for Fundamental Physics}},\ }\href {https://doi.org/10.1146/annurev-nucl-101916-123130} {\bibfield  {journal} {\bibinfo  {journal} {Ann. Rev. Nucl. Part. Sci.}\ }\textbf {\bibinfo {volume} {67}},\ \bibinfo {pages} {161} (\bibinfo {year} {2017})}\BibitemShut {NoStop}%
\bibitem [{\citenamefont {Pattavina}\ \emph {et~al.}(2019)\citenamefont {Pattavina} \emph {et~al.}}]{Pattavina:2019pxw}%
  \BibitemOpen
  \bibfield  {author} {\bibinfo {author} {\bibfnamefont {L.}~\bibnamefont {Pattavina}} \emph {et~al.},\ }\bibfield  {title} {\bibinfo {title} {{Radiopurity of an archeological Roman Lead cryogenic detector}},\ }\href {https://doi.org/10.1140/epja/i2019-12809-0} {\bibfield  {journal} {\bibinfo  {journal} {Eur. Phys. J.}\ }\textbf {\bibinfo {volume} {A 55}},\ \bibinfo {pages} {127} (\bibinfo {year} {2019})},\ \Eprint {https://arxiv.org/abs/1904.04040} {arXiv:1904.04040 [physics.ins-det]} \BibitemShut {NoStop}%
\bibitem [{\citenamefont {Drukier}\ and\ \citenamefont {Stodolsky}(1984)}]{Drukier:1983gj}%
  \BibitemOpen
  \bibfield  {author} {\bibinfo {author} {\bibfnamefont {A.}~\bibnamefont {Drukier}}\ and\ \bibinfo {author} {\bibfnamefont {L.}~\bibnamefont {Stodolsky}},\ }\bibfield  {title} {\bibinfo {title} {{Principles and Applications of a Neutral Current Detector for Neutrino Physics and Astronomy}},\ }\href {https://doi.org/10.1103/PhysRevD.30.2295} {\bibfield  {journal} {\bibinfo  {journal} {Phys. Rev.}\ }\textbf {\bibinfo {volume} {D 30}},\ \bibinfo {pages} {2295} (\bibinfo {year} {1984})}\BibitemShut {NoStop}%
\bibitem [{\citenamefont {Suliga}\ \emph {et~al.}(2022)\citenamefont {Suliga}, \citenamefont {Beacom},\ and\ \citenamefont {Tamborra}}]{Suliga_DSNB}%
  \BibitemOpen
  \bibfield  {author} {\bibinfo {author} {\bibfnamefont {A.~M.}\ \bibnamefont {Suliga}}, \bibinfo {author} {\bibfnamefont {J.~F.}\ \bibnamefont {Beacom}},\ and\ \bibinfo {author} {\bibfnamefont {I.}~\bibnamefont {Tamborra}},\ }\bibfield  {title} {\bibinfo {title} {Towards probing the diffuse supernova neutrino background in all flavors},\ }\href {https://doi.org/10.1103/PhysRevD.105.043008} {\bibfield  {journal} {\bibinfo  {journal} {Phys. Rev. D}\ }\textbf {\bibinfo {volume} {105}},\ \bibinfo {pages} {043008} (\bibinfo {year} {2022})}\BibitemShut {NoStop}%
\bibitem [{\citenamefont {Huang}\ and\ \citenamefont {Chen}(2022)}]{SN_skin}%
  \BibitemOpen
  \bibfield  {author} {\bibinfo {author} {\bibfnamefont {X.-R.}\ \bibnamefont {Huang}}\ and\ \bibinfo {author} {\bibfnamefont {L.-W.}\ \bibnamefont {Chen}},\ }\bibfield  {title} {\bibinfo {title} {{Supernova neutrinos as a precise probe of nuclear neutron skin}},\ }\href {https://doi.org/10.1103/PhysRevD.106.123034} {\bibfield  {journal} {\bibinfo  {journal} {Phys. Rev. D}\ }\textbf {\bibinfo {volume} {106}},\ \bibinfo {pages} {123034} (\bibinfo {year} {2022})},\ \Eprint {https://arxiv.org/abs/2210.04534} {arXiv:2210.04534 [nucl-th]} \BibitemShut {NoStop}%
\bibitem [{\citenamefont {Sen}(2024)}]{Sen:2024fxa}%
  \BibitemOpen
  \bibfield  {author} {\bibinfo {author} {\bibfnamefont {M.}~\bibnamefont {Sen}},\ }\bibfield  {title} {\bibinfo {title} {{Supernova Neutrinos: Flavour Conversion Mechanisms and New Physics Scenarios}},\ }\href {https://doi.org/10.3390/universe10060238} {\bibfield  {journal} {\bibinfo  {journal} {Universe}\ }\textbf {\bibinfo {volume} {10}},\ \bibinfo {pages} {238} (\bibinfo {year} {2024})},\ \Eprint {https://arxiv.org/abs/2405.20432} {arXiv:2405.20432 [hep-ph]} \BibitemShut {NoStop}%
\bibitem [{\citenamefont {Catena}\ and\ \citenamefont {Covi}(2014)}]{Catena:2013pka}%
  \BibitemOpen
  \bibfield  {author} {\bibinfo {author} {\bibfnamefont {R.}~\bibnamefont {Catena}}\ and\ \bibinfo {author} {\bibfnamefont {L.}~\bibnamefont {Covi}},\ }\bibfield  {title} {\bibinfo {title} {{SUSY dark matter(s)}},\ }\href {https://doi.org/10.1140/epjc/s10052-013-2703-4} {\bibfield  {journal} {\bibinfo  {journal} {Eur. Phys. J. C}\ }\textbf {\bibinfo {volume} {74}},\ \bibinfo {pages} {2703} (\bibinfo {year} {2014})},\ \Eprint {https://arxiv.org/abs/1310.4776} {arXiv:1310.4776 [hep-ph]} \BibitemShut {NoStop}%
\bibitem [{\citenamefont {Silveira}\ and\ \citenamefont {Zee}(1985)}]{SILVEIRA1985136}%
  \BibitemOpen
  \bibfield  {author} {\bibinfo {author} {\bibfnamefont {V.}~\bibnamefont {Silveira}}\ and\ \bibinfo {author} {\bibfnamefont {A.}~\bibnamefont {Zee}},\ }\bibfield  {title} {\bibinfo {title} {Scalar phantoms},\ }\href {https://doi.org/10.1016/0370-2693(85)90624-0} {\bibfield  {journal} {\bibinfo  {journal} {Phys. Lett. B}\ }\textbf {\bibinfo {volume} {161}},\ \bibinfo {pages} {136} (\bibinfo {year} {1985})}\BibitemShut {NoStop}%
\bibitem [{\citenamefont {Kanemura}\ \emph {et~al.}(2010)\citenamefont {Kanemura}, \citenamefont {Matsumoto}, \citenamefont {Nabeshima},\ and\ \citenamefont {Okada}}]{Kanemura:2010sh}%
  \BibitemOpen
  \bibfield  {author} {\bibinfo {author} {\bibfnamefont {S.}~\bibnamefont {Kanemura}}, \bibinfo {author} {\bibfnamefont {S.}~\bibnamefont {Matsumoto}}, \bibinfo {author} {\bibfnamefont {T.}~\bibnamefont {Nabeshima}},\ and\ \bibinfo {author} {\bibfnamefont {N.}~\bibnamefont {Okada}},\ }\bibfield  {title} {\bibinfo {title} {{Can WIMP Dark Matter overcome the Nightmare Scenario?}},\ }\href {https://doi.org/10.1103/PhysRevD.82.055026} {\bibfield  {journal} {\bibinfo  {journal} {Phys. Rev. D}\ }\textbf {\bibinfo {volume} {82}},\ \bibinfo {pages} {055026} (\bibinfo {year} {2010})},\ \Eprint {https://arxiv.org/abs/1005.5651} {arXiv:1005.5651 [hep-ph]} \BibitemShut {NoStop}%
\bibitem [{\citenamefont {Arbey}\ and\ \citenamefont {Mahmoudi}(2021)}]{ARBEY2021103865}%
  \BibitemOpen
  \bibfield  {author} {\bibinfo {author} {\bibfnamefont {A.}~\bibnamefont {Arbey}}\ and\ \bibinfo {author} {\bibfnamefont {F.}~\bibnamefont {Mahmoudi}},\ }\bibfield  {title} {\bibinfo {title} {Dark matter and the early universe: A review},\ }\href {https://doi.org/10.1016/j.ppnp.2021.103865} {\bibfield  {journal} {\bibinfo  {journal} {Prog. Part. Nucl. Phys.}\ }\textbf {\bibinfo {volume} {119}},\ \bibinfo {pages} {103865} (\bibinfo {year} {2021})}\BibitemShut {NoStop}%
\bibitem [{\citenamefont {Nussinov}(1985)}]{NUSSINOV198555}%
  \BibitemOpen
  \bibfield  {author} {\bibinfo {author} {\bibfnamefont {S.}~\bibnamefont {Nussinov}},\ }\bibfield  {title} {\bibinfo {title} {Technocosmology — could a technibaryon excess provide a “natural” missing mass candidate?},\ }\href {https://doi.org/10.1016/0370-2693(85)90689-6} {\bibfield  {journal} {\bibinfo  {journal} {Phys. Lett. B}\ }\textbf {\bibinfo {volume} {165}},\ \bibinfo {pages} {55} (\bibinfo {year} {1985})}\BibitemShut {NoStop}%
\bibitem [{\citenamefont {Billard}\ \emph {et~al.}(2014{\natexlab{a}})\citenamefont {Billard}, \citenamefont {Figueroa-Feliciano},\ and\ \citenamefont {Strigari}}]{Billard}%
  \BibitemOpen
  \bibfield  {author} {\bibinfo {author} {\bibfnamefont {J.}~\bibnamefont {Billard}}, \bibinfo {author} {\bibfnamefont {E.}~\bibnamefont {Figueroa-Feliciano}},\ and\ \bibinfo {author} {\bibfnamefont {L.}~\bibnamefont {Strigari}},\ }\bibfield  {title} {\bibinfo {title} {Implication of neutrino backgrounds on the reach of next generation dark matter direct detection experiments},\ }\href {https://doi.org/10.1103/PhysRevD.89.023524} {\bibfield  {journal} {\bibinfo  {journal} {Phys. Rev. D}\ }\textbf {\bibinfo {volume} {89}},\ \bibinfo {pages} {023524} (\bibinfo {year} {2014}{\natexlab{a}})}\BibitemShut {NoStop}%
\bibitem [{\citenamefont {Sadoulet}(2024)}]{Sadoulet}%
  \BibitemOpen
  \bibfield  {author} {\bibinfo {author} {\bibfnamefont {B.}~\bibnamefont {Sadoulet}},\ }\bibfield  {title} {\bibinfo {title} {Forty years of dark matter searches},\ }\href {https://doi.org/10.1016/j.nuclphysb.2024.116509} {\bibfield  {journal} {\bibinfo  {journal} {Nucl. Phys. B}\ }\textbf {\bibinfo {volume} {1003}},\ \bibinfo {pages} {116509} (\bibinfo {year} {2024})}\BibitemShut {NoStop}%
\bibitem [{\citenamefont {Freedman}(1974)}]{Freedman:1973yd}%
  \BibitemOpen
  \bibfield  {author} {\bibinfo {author} {\bibfnamefont {D.~Z.}\ \bibnamefont {Freedman}},\ }\bibfield  {title} {\bibinfo {title} {{Coherent Neutrino Nucleus Scattering as a Probe of the Weak Neutral Current}},\ }\href {https://doi.org/10.1103/PhysRevD.9.1389} {\bibfield  {journal} {\bibinfo  {journal} {Phys. Rev. D}\ }\textbf {\bibinfo {volume} {9}},\ \bibinfo {pages} {1389} (\bibinfo {year} {1974})}\BibitemShut {NoStop}%
\bibitem [{\citenamefont {Scholberg}(2012)}]{Scholberg:2012id}%
  \BibitemOpen
  \bibfield  {author} {\bibinfo {author} {\bibfnamefont {K.}~\bibnamefont {Scholberg}},\ }\bibfield  {title} {\bibinfo {title} {{Supernova Neutrino Detection}},\ }\href {https://doi.org/10.1146/annurev-nucl-102711-095006} {\bibfield  {journal} {\bibinfo  {journal} {Ann. Rev. Nucl. Part. Sci.}\ }\textbf {\bibinfo {volume} {62}},\ \bibinfo {pages} {81} (\bibinfo {year} {2012})},\ \Eprint {https://arxiv.org/abs/1205.6003} {arXiv:1205.6003 [astro-ph.IM]} \BibitemShut {NoStop}%
\bibitem [{Note1()}]{Note1}%
  \BibitemOpen
  \bibinfo {note} {($4 \sin ^2 \theta _W -1) \approx 0.05$.}\BibitemShut {Stop}%
\bibitem [{\citenamefont {Nosengo}(2010)}]{Nosengo}%
  \BibitemOpen
  \bibfield  {author} {\bibinfo {author} {\bibfnamefont {N.}~\bibnamefont {Nosengo}},\ }\bibfield  {title} {\bibinfo {title} {Roman ingots to shield particle detector},\ }\bibfield  {journal} {\bibinfo  {journal} {Nature}\ }\href {https://doi.org/10.1038/news.2010.186} {10.1038/news.2010.186} (\bibinfo {year} {2010})\BibitemShut {NoStop}%
\bibitem [{\citenamefont {Misiaszek}\ and\ \citenamefont {Rossi}(2024)}]{DM_review2024}%
  \BibitemOpen
  \bibfield  {author} {\bibinfo {author} {\bibfnamefont {M.}~\bibnamefont {Misiaszek}}\ and\ \bibinfo {author} {\bibfnamefont {N.}~\bibnamefont {Rossi}},\ }\bibfield  {title} {\bibinfo {title} {Direct detection of dark matter: A critical review},\ }\href {https://doi.org/10.3390/sym16020201} {\bibfield  {journal} {\bibinfo  {journal} {Symmetry}\ }\textbf {\bibinfo {volume} {16}},\ \bibinfo {pages} {201} (\bibinfo {year} {2024})}\BibitemShut {NoStop}%
\bibitem [{\citenamefont {Freese}\ \emph {et~al.}(2013)\citenamefont {Freese}, \citenamefont {Lisanti},\ and\ \citenamefont {Savage}}]{Review_modulation}%
  \BibitemOpen
  \bibfield  {author} {\bibinfo {author} {\bibfnamefont {K.}~\bibnamefont {Freese}}, \bibinfo {author} {\bibfnamefont {M.}~\bibnamefont {Lisanti}},\ and\ \bibinfo {author} {\bibfnamefont {C.}~\bibnamefont {Savage}},\ }\bibfield  {title} {\bibinfo {title} {Colloquium: Annual modulation of dark matter},\ }\href {https://doi.org/10.1103/RevModPhys.85.1561} {\bibfield  {journal} {\bibinfo  {journal} {Rev. Mod. Phys.}\ }\textbf {\bibinfo {volume} {85}},\ \bibinfo {pages} {1561} (\bibinfo {year} {2013})}\BibitemShut {NoStop}%
\bibitem [{\citenamefont {Beeman}\ \emph {et~al.}(2022)\citenamefont {Beeman} \emph {et~al.}}]{kg-scale}%
  \BibitemOpen
  \bibfield  {author} {\bibinfo {author} {\bibfnamefont {J.~W.}\ \bibnamefont {Beeman}} \emph {et~al.} (\bibinfo {collaboration} {{RES-NOVA}}),\ }\bibfield  {title} {\bibinfo {title} {{Radiopurity of a kg-scale $PbWO_4$ cryogenic detector produced from archaeological Pb for the RES-NOVA experiment}},\ }\href {https://doi.org/10.1140/epjc/s10052-022-10656-8} {\bibfield  {journal} {\bibinfo  {journal} {Eur. Phys. J. C}\ }\textbf {\bibinfo {volume} {82}},\ \bibinfo {pages} {692} (\bibinfo {year} {2022})}\BibitemShut {NoStop}%
\bibitem [{\citenamefont {Lewin}\ and\ \citenamefont {Smith}(1996)}]{Lewin:1995rx}%
  \BibitemOpen
  \bibfield  {author} {\bibinfo {author} {\bibfnamefont {J.~D.}\ \bibnamefont {Lewin}}\ and\ \bibinfo {author} {\bibfnamefont {P.~F.}\ \bibnamefont {Smith}},\ }\bibfield  {title} {\bibinfo {title} {{Review of mathematics, numerical factors, and corrections for dark matter experiments based on elastic nuclear recoil}},\ }\href {https://doi.org/10.1016/S0927-6505(96)00047-3} {\bibfield  {journal} {\bibinfo  {journal} {Astropart. Phys.}\ }\textbf {\bibinfo {volume} {6}},\ \bibinfo {pages} {87} (\bibinfo {year} {1996})}\BibitemShut {NoStop}%
\bibitem [{\citenamefont {Schumann}(2019)}]{Schumann:2019eaa}%
  \BibitemOpen
  \bibfield  {author} {\bibinfo {author} {\bibfnamefont {M.}~\bibnamefont {Schumann}},\ }\bibfield  {title} {\bibinfo {title} {{Direct Detection of WIMP Dark Matter: Concepts and Status}},\ }\href {https://doi.org/10.1088/1361-6471/ab2ea5} {\bibfield  {journal} {\bibinfo  {journal} {J. Phys. G}\ }\textbf {\bibinfo {volume} {46}},\ \bibinfo {pages} {103003} (\bibinfo {year} {2019})},\ \Eprint {https://arxiv.org/abs/1903.03026} {arXiv:1903.03026 [astro-ph.CO]} \BibitemShut {NoStop}%
\bibitem [{\citenamefont {{Norman, E. and Holden, N.E.}}(2018)}]{IUPAP}%
  \BibitemOpen
  \bibfield  {author} {\bibinfo {author} {\bibnamefont {{Norman, E. and Holden, N.E.}}},\ }\bibfield  {title} {\bibinfo {title} {{IUPAC} periodic table of the elements and isotopes ({IPTEI}) for the education community ({IUPAC} technical report)},\ }\href {https://doi.org/10.1515/pac-2015-0703} {\bibfield  {journal} {\bibinfo  {journal} {Pure Appl. Chem.}\ }\textbf {\bibinfo {volume} {90}},\ \bibinfo {pages} {1833} (\bibinfo {year} {2018})}\BibitemShut {NoStop}%
\bibitem [{\citenamefont {Kondev}\ \emph {et~al.}(2021)\citenamefont {Kondev}, \citenamefont {Wang}, \citenamefont {Huang}, \citenamefont {Naimi},\ and\ \citenamefont {Audi}}]{Kondev_2021}%
  \BibitemOpen
  \bibfield  {author} {\bibinfo {author} {\bibfnamefont {F.}~\bibnamefont {Kondev}}, \bibinfo {author} {\bibfnamefont {M.}~\bibnamefont {Wang}}, \bibinfo {author} {\bibfnamefont {W.}~\bibnamefont {Huang}}, \bibinfo {author} {\bibfnamefont {S.}~\bibnamefont {Naimi}},\ and\ \bibinfo {author} {\bibfnamefont {G.}~\bibnamefont {Audi}},\ }\bibfield  {title} {\bibinfo {title} {The {NUBASE2020} evaluation of nuclear physics properties},\ }\href {https://doi.org/10.1088/1674-1137/abddae} {\bibfield  {journal} {\bibinfo  {journal} {Chin. Phys. C}\ }\textbf {\bibinfo {volume} {45}},\ \bibinfo {pages} {030001} (\bibinfo {year} {2021})}\BibitemShut {NoStop}%
\bibitem [{\citenamefont {Engel}\ \emph {et~al.}(1992)\citenamefont {Engel}, \citenamefont {Pittel},\ and\ \citenamefont {Vogel}}]{Engel:1992bf}%
  \BibitemOpen
  \bibfield  {author} {\bibinfo {author} {\bibfnamefont {J.}~\bibnamefont {Engel}}, \bibinfo {author} {\bibfnamefont {S.}~\bibnamefont {Pittel}},\ and\ \bibinfo {author} {\bibfnamefont {P.}~\bibnamefont {Vogel}},\ }\bibfield  {title} {\bibinfo {title} {{Nuclear physics of dark matter detection}},\ }\href {https://doi.org/10.1142/S0218301392000023} {\bibfield  {journal} {\bibinfo  {journal} {Int. J. Mod. Phys. E}\ }\textbf {\bibinfo {volume} {1}},\ \bibinfo {pages} {1} (\bibinfo {year} {1992})}\BibitemShut {NoStop}%
\bibitem [{\citenamefont {Bednyakov}\ and\ \citenamefont {Simkovic}(2006)}]{Bednyakov:2006ux}%
  \BibitemOpen
  \bibfield  {author} {\bibinfo {author} {\bibfnamefont {V.~A.}\ \bibnamefont {Bednyakov}}\ and\ \bibinfo {author} {\bibfnamefont {F.}~\bibnamefont {Simkovic}},\ }\bibfield  {title} {\bibinfo {title} {{Nuclear spin structure in dark matter search: The Finite momentum transfer limit}},\ }\href {https://doi.org/10.1134/S1063779606070057} {\bibfield  {journal} {\bibinfo  {journal} {Phys. Part. Nucl.}\ }\textbf {\bibinfo {volume} {37}},\ \bibinfo {pages} {S106} (\bibinfo {year} {2006})},\ \Eprint {https://arxiv.org/abs/hep-ph/0608097} {arXiv:hep-ph/0608097} \BibitemShut {NoStop}%
\bibitem [{\citenamefont {Kosmas}\ and\ \citenamefont {Vergados}(1997)}]{Kosmas:1997jm}%
  \BibitemOpen
  \bibfield  {author} {\bibinfo {author} {\bibfnamefont {T.~S.}\ \bibnamefont {Kosmas}}\ and\ \bibinfo {author} {\bibfnamefont {J.~D.}\ \bibnamefont {Vergados}},\ }\bibfield  {title} {\bibinfo {title} {{Cold dark matter in SUSY theories. The Role of nuclear form-factors and the folding with the LSP velocity}},\ }\href {https://doi.org/10.1103/PhysRevD.55.1752} {\bibfield  {journal} {\bibinfo  {journal} {Phys. Rev. D}\ }\textbf {\bibinfo {volume} {55}},\ \bibinfo {pages} {1752} (\bibinfo {year} {1997})},\ \Eprint {https://arxiv.org/abs/hep-ph/9701205} {arXiv:hep-ph/9701205} \BibitemShut {NoStop}%
\bibitem [{\citenamefont {Bellini}\ \emph {et~al.}(2012)\citenamefont {Bellini} \emph {et~al.}}]{G.Bellini_2012}%
  \BibitemOpen
  \bibfield  {author} {\bibinfo {author} {\bibfnamefont {G.}~\bibnamefont {Bellini}} \emph {et~al.},\ }\bibfield  {title} {\bibinfo {title} {Cosmic-muon flux and annual modulation in borexino at 3800 m water-equivalent depth},\ }\href {https://doi.org/10.1088/1475-7516/2012/05/015} {\bibfield  {journal} {\bibinfo  {journal} {J. Cosmol. Astropart. Phys.}\ }\textbf {\bibinfo {volume} {2012}}\bibinfo  {number} { (05)},\ \bibinfo {pages} {015}}\BibitemShut {NoStop}%
\bibitem [{\citenamefont {Ferreiro~Iachellini}\ \emph {et~al.}(2022)\citenamefont {Ferreiro~Iachellini} \emph {et~al.}}]{FerreiroIachellini:2021qgu}%
  \BibitemOpen
\bibfield  {number} {  }\bibfield  {author} {\bibinfo {author} {\bibfnamefont {N.}~\bibnamefont {Ferreiro~Iachellini}} \emph {et~al.},\ }\bibfield  {title} {\bibinfo {title} {{Operation of an Archaeological Lead PbWO$_4$ Crystal to Search for Neutrinos from Astrophysical Sources with a Transition Edge Sensor}},\ }\href {https://doi.org/10.1007/s10909-022-02823-8} {\bibfield  {journal} {\bibinfo  {journal} {J. Low Temp. Phys.}\ }\textbf {\bibinfo {volume} {209}},\ \bibinfo {pages} {872} (\bibinfo {year} {2022})},\ \Eprint {https://arxiv.org/abs/2111.07638} {arXiv:2111.07638 [physics.ins-det]} \BibitemShut {NoStop}%
\bibitem [{\citenamefont {Pr{\"o}bst}\ \emph {et~al.}(1995)\citenamefont {Pr{\"o}bst}, \citenamefont {Frank}, \citenamefont {Cooper}, \citenamefont {Colling}, \citenamefont {Dummer}, \citenamefont {Ferger}, \citenamefont {Forster}, \citenamefont {Nucciotti}, \citenamefont {Seidel},\ and\ \citenamefont {Stodolsky}}]{Franz}%
  \BibitemOpen
  \bibfield  {author} {\bibinfo {author} {\bibfnamefont {F.}~\bibnamefont {Pr{\"o}bst}}, \bibinfo {author} {\bibfnamefont {M.}~\bibnamefont {Frank}}, \bibinfo {author} {\bibfnamefont {S.}~\bibnamefont {Cooper}}, \bibinfo {author} {\bibfnamefont {P.}~\bibnamefont {Colling}}, \bibinfo {author} {\bibfnamefont {D.}~\bibnamefont {Dummer}}, \bibinfo {author} {\bibfnamefont {P.}~\bibnamefont {Ferger}}, \bibinfo {author} {\bibfnamefont {G.}~\bibnamefont {Forster}}, \bibinfo {author} {\bibfnamefont {A.}~\bibnamefont {Nucciotti}}, \bibinfo {author} {\bibfnamefont {W.}~\bibnamefont {Seidel}},\ and\ \bibinfo {author} {\bibfnamefont {L.}~\bibnamefont {Stodolsky}},\ }\bibfield  {title} {\bibinfo {title} {Model for cryogenic particle detectors with superconducting phase transition thermometers},\ }\href {https://doi.org/10.1007/BF00753837} {\bibfield  {journal} {\bibinfo  {journal} {J. Low Temp. Phys.}\ }\textbf {\bibinfo {volume} {100}},\ \bibinfo {pages} {69} (\bibinfo {year} {1995})}\BibitemShut {NoStop}%
\bibitem [{\citenamefont {Strauss}\ \emph {et~al.}(2017)\citenamefont {Strauss} \emph {et~al.}}]{Strauss:2017cam}%
  \BibitemOpen
  \bibfield  {author} {\bibinfo {author} {\bibfnamefont {R.}~\bibnamefont {Strauss}} \emph {et~al.},\ }\bibfield  {title} {\bibinfo {title} {{Gram-scale cryogenic calorimeters for rare-event searches}},\ }\href {https://doi.org/10.1103/PhysRevD.96.022009} {\bibfield  {journal} {\bibinfo  {journal} {Phys. Rev. D}\ }\textbf {\bibinfo {volume} {96}},\ \bibinfo {pages} {022009} (\bibinfo {year} {2017})},\ \Eprint {https://arxiv.org/abs/1704.04317} {arXiv:1704.04317 [physics.ins-det]} \BibitemShut {NoStop}%
\bibitem [{\citenamefont {Beeman}\ \emph {et~al.}(2013{\natexlab{a}})\citenamefont {Beeman} \emph {et~al.}}]{Beeman2013}%
  \BibitemOpen
  \bibfield  {author} {\bibinfo {author} {\bibfnamefont {J.}~\bibnamefont {Beeman}} \emph {et~al.},\ }\bibfield  {title} {\bibinfo {title} {Characterization of bolometric light detectors for rare event searches},\ }\href {https://doi.org/10.1088/1748-0221/8/07/P07021} {\bibfield  {journal} {\bibinfo  {journal} {J. Instrum.}\ }\textbf {\bibinfo {volume} {8}}\bibinfo  {number} { (7)}}\BibitemShut {NoStop}%
\bibitem [{\citenamefont {Beeman}\ \emph {et~al.}(2013{\natexlab{b}})\citenamefont {Beeman} \emph {et~al.}}]{Beeman:2012wz}%
  \BibitemOpen
\bibfield  {number} {  }\bibfield  {author} {\bibinfo {author} {\bibfnamefont {J.~W.}\ \bibnamefont {Beeman}} \emph {et~al.},\ }\bibfield  {title} {\bibinfo {title} {{New experimental limits on the alpha decays of lead isotopes}},\ }\href {https://doi.org/10.1140/epja/i2013-13050-7} {\bibfield  {journal} {\bibinfo  {journal} {Eur. Phys. J.}\ }\textbf {\bibinfo {volume} {A 49}},\ \bibinfo {pages} {50} (\bibinfo {year} {2013}{\natexlab{b}})},\ \Eprint {https://arxiv.org/abs/1212.2422} {arXiv:1212.2422 [nucl-ex]} \BibitemShut {NoStop}%
\bibitem [{\citenamefont {Pattavina}\ \emph {et~al.}(2021)\citenamefont {Pattavina} \emph {et~al.}}]{Pattavina_2021}%
  \BibitemOpen
  \bibfield  {author} {\bibinfo {author} {\bibfnamefont {L.}~\bibnamefont {Pattavina}} \emph {et~al.},\ }\bibfield  {title} {\bibinfo {title} {{RES-NOVA} sensitivity to core-collapse and failed core-collapse supernova neutrinos},\ }\href {https://doi.org/10.1088/1475-7516/2021/10/064} {\bibfield  {journal} {\bibinfo  {journal} {J. Cosmol. Astropart. Phys.}\ }\textbf {\bibinfo {volume} {2021}}\bibinfo  {number} { (10)},\ \bibinfo {pages} {064}}\BibitemShut {NoStop}%
\bibitem [{\citenamefont {Clemenza}\ \emph {et~al.}(2011)\citenamefont {Clemenza}, \citenamefont {Maiano}, \citenamefont {Pattavina},\ and\ \citenamefont {Previtali}}]{Clemenza:2011zz}%
  \BibitemOpen
\bibfield  {number} {  }\bibfield  {author} {\bibinfo {author} {\bibfnamefont {M.}~\bibnamefont {Clemenza}}, \bibinfo {author} {\bibfnamefont {C.}~\bibnamefont {Maiano}}, \bibinfo {author} {\bibfnamefont {L.}~\bibnamefont {Pattavina}},\ and\ \bibinfo {author} {\bibfnamefont {E.}~\bibnamefont {Previtali}},\ }\bibfield  {title} {\bibinfo {title} {{Radon-induced surface contaminations in low background experiments}},\ }\href {https://doi.org/10.1140/epjc/s10052-011-1805-0} {\bibfield  {journal} {\bibinfo  {journal} {Eur. Phys. J. C}\ }\textbf {\bibinfo {volume} {71}},\ \bibinfo {pages} {1805} (\bibinfo {year} {2011})}\BibitemShut {NoStop}%
\bibitem [{\citenamefont {Belli}\ \emph {et~al.}(2020)\citenamefont {Belli} \emph {et~al.}}]{Belli:2020qqc}%
  \BibitemOpen
  \bibfield  {author} {\bibinfo {author} {\bibfnamefont {P.}~\bibnamefont {Belli}} \emph {et~al.},\ }\bibfield  {title} {\bibinfo {title} {{Search for Double Beta Decay of $^{106}$Cd with an Enriched $^{106}$CdWO$_4$ Crystal Scintillator in Coincidence with CdWO$_4$ Scintillation Counters}},\ }\href {https://doi.org/10.3390/universe6100182} {\bibfield  {journal} {\bibinfo  {journal} {Universe}\ }\textbf {\bibinfo {volume} {6}},\ \bibinfo {pages} {182} (\bibinfo {year} {2020})},\ \Eprint {https://arxiv.org/abs/2010.08749} {arXiv:2010.08749 [nucl-ex]} \BibitemShut {NoStop}%
\bibitem [{\citenamefont {Adams}\ \emph {et~al.}(2024)\citenamefont {Adams} \emph {et~al.}}]{Data-driven}%
  \BibitemOpen
  \bibfield  {author} {\bibinfo {author} {\bibfnamefont {D.~Q.}\ \bibnamefont {Adams}} \emph {et~al.} (\bibinfo {collaboration} {{CUORE}}),\ }\bibfield  {title} {\bibinfo {title} {Data-driven background model for the {CUORE} experiment},\ }\href {https://doi.org/10.1103/PhysRevD.110.052003} {\bibfield  {journal} {\bibinfo  {journal} {Phys. Rev. D}\ }\textbf {\bibinfo {volume} {110}},\ \bibinfo {pages} {052003} (\bibinfo {year} {2024})}\BibitemShut {NoStop}%
\bibitem [{\citenamefont {Aprile}(2010)}]{Aprile:2010zz}%
  \BibitemOpen
  \bibfield  {author} {\bibinfo {author} {\bibfnamefont {E.}~\bibnamefont {Aprile}} (\bibinfo {collaboration} {{XENON100}}),\ }\bibfield  {title} {\bibinfo {title} {{The XENON100 dark matter experiment at LNGS: Status and sensitivity}},\ }\href {https://doi.org/10.1088/1742-6596/203/1/012005} {\bibfield  {journal} {\bibinfo  {journal} {J. Phys. Conf. Ser.}\ }\textbf {\bibinfo {volume} {203}},\ \bibinfo {pages} {012005} (\bibinfo {year} {2010})}\BibitemShut {NoStop}%
\bibitem [{\citenamefont {Wulandari}\ \emph {et~al.}(2004)\citenamefont {Wulandari}, \citenamefont {Jochum}, \citenamefont {Rau},\ and\ \citenamefont {von Feilitzsch}}]{Wulandari:2003cr}%
  \BibitemOpen
  \bibfield  {author} {\bibinfo {author} {\bibfnamefont {H.}~\bibnamefont {Wulandari}}, \bibinfo {author} {\bibfnamefont {J.}~\bibnamefont {Jochum}}, \bibinfo {author} {\bibfnamefont {W.}~\bibnamefont {Rau}},\ and\ \bibinfo {author} {\bibfnamefont {F.}~\bibnamefont {von Feilitzsch}},\ }\bibfield  {title} {\bibinfo {title} {{Neutron flux underground revisited}},\ }\href {https://doi.org/10.1016/j.astropartphys.2004.07.005} {\bibfield  {journal} {\bibinfo  {journal} {Astropart. Phys.}\ }\textbf {\bibinfo {volume} {22}},\ \bibinfo {pages} {313} (\bibinfo {year} {2004})},\ \Eprint {https://arxiv.org/abs/hep-ex/0312050} {arXiv:hep-ex/0312050} \BibitemShut {NoStop}%
\bibitem [{\citenamefont {Malczewski}\ \emph {et~al.}(2013)\citenamefont {Malczewski}, \citenamefont {Kisiel},\ and\ \citenamefont {Dorda}}]{LNGS-gamma}%
  \BibitemOpen
  \bibfield  {author} {\bibinfo {author} {\bibfnamefont {D.}~\bibnamefont {Malczewski}}, \bibinfo {author} {\bibfnamefont {J.}~\bibnamefont {Kisiel}},\ and\ \bibinfo {author} {\bibfnamefont {J.}~\bibnamefont {Dorda}},\ }\bibfield  {title} {\bibinfo {title} {Gamma background measurements in the {Gran Sasso National Laboratory}},\ }\href {https://doi.org/10.1007/s10967-012-1990-9} {\bibfield  {journal} {\bibinfo  {journal} {J. Radioanal. Nucl. Chem.}\ }\textbf {\bibinfo {volume} {295}},\ \bibinfo {pages} {749} (\bibinfo {year} {2013})}\BibitemShut {NoStop}%
\bibitem [{\citenamefont {Agostinelli}\ \emph {et~al.}(2003)\citenamefont {Agostinelli} \emph {et~al.}}]{Agostinelli:2002hh}%
  \BibitemOpen
  \bibfield  {author} {\bibinfo {author} {\bibfnamefont {S.}~\bibnamefont {Agostinelli}} \emph {et~al.} (\bibinfo {collaboration} {GEANT4}),\ }\bibfield  {title} {\bibinfo {title} {{GEANT4--a simulation toolkit}},\ }\href {https://doi.org/10.1016/S0168-9002(03)01368-8} {\bibfield  {journal} {\bibinfo  {journal} {Nucl. Instrum. Meth. A}\ }\textbf {\bibinfo {volume} {506}},\ \bibinfo {pages} {250} (\bibinfo {year} {2003})}\BibitemShut {NoStop}%
\bibitem [{Note2()}]{Note2}%
  \BibitemOpen
  \bibinfo {note} {Equivalent to 0.086~c/keV/kg/d.}\BibitemShut {Stop}%
\bibitem [{\citenamefont {Cowan}\ \emph {et~al.}(2011)\citenamefont {Cowan}, \citenamefont {Cranmer}, \citenamefont {Gross},\ and\ \citenamefont {Vitells}}]{Cowan:2010js}%
  \BibitemOpen
  \bibfield  {author} {\bibinfo {author} {\bibfnamefont {G.}~\bibnamefont {Cowan}}, \bibinfo {author} {\bibfnamefont {K.}~\bibnamefont {Cranmer}}, \bibinfo {author} {\bibfnamefont {E.}~\bibnamefont {Gross}},\ and\ \bibinfo {author} {\bibfnamefont {O.}~\bibnamefont {Vitells}},\ }\bibfield  {title} {\bibinfo {title} {{Asymptotic formulae for likelihood-based tests of new physics}},\ }\href {https://doi.org/10.1140/epjc/s10052-011-1554-0} {\bibfield  {journal} {\bibinfo  {journal} {Eur. Phys. J. C}\ }\textbf {\bibinfo {volume} {71}},\ \bibinfo {pages} {1554} (\bibinfo {year} {2011})},\ \Eprint {https://arxiv.org/abs/1007.1727} {arXiv:1007.1727 [physics.data-an]} \BibitemShut {NoStop}%
\bibitem [{\citenamefont {Heinrich}\ \emph {et~al.}()\citenamefont {Heinrich}, \citenamefont {Feickert},\ and\ \citenamefont {Stark}}]{pyhf}%
  \BibitemOpen
  \bibfield  {author} {\bibinfo {author} {\bibfnamefont {L.}~\bibnamefont {Heinrich}}, \bibinfo {author} {\bibfnamefont {M.}~\bibnamefont {Feickert}},\ and\ \bibinfo {author} {\bibfnamefont {G.}~\bibnamefont {Stark}},\ }\href {https://doi.org/10.5281/zenodo.1169739} {\bibinfo {title} {{pyhf: v0.7.6}}},\ \bibinfo {note} {https://github.com/scikit-hep/pyhf/releases/tag/v0.7.6}\BibitemShut {NoStop}%
\bibitem [{\citenamefont {Heinrich}\ \emph {et~al.}(2021)\citenamefont {Heinrich}, \citenamefont {Feickert}, \citenamefont {Stark},\ and\ \citenamefont {Cranmer}}]{pyhf_joss}%
  \BibitemOpen
  \bibfield  {author} {\bibinfo {author} {\bibfnamefont {L.}~\bibnamefont {Heinrich}}, \bibinfo {author} {\bibfnamefont {M.}~\bibnamefont {Feickert}}, \bibinfo {author} {\bibfnamefont {G.}~\bibnamefont {Stark}},\ and\ \bibinfo {author} {\bibfnamefont {K.}~\bibnamefont {Cranmer}},\ }\bibfield  {title} {\bibinfo {title} {pyhf: pure-python implementation of histfactory statistical models},\ }\href {https://doi.org/10.21105/joss.02823} {\bibfield  {journal} {\bibinfo  {journal} {J. Open Source Softw.}\ }\textbf {\bibinfo {volume} {6}},\ \bibinfo {pages} {2823} (\bibinfo {year} {2021})}\BibitemShut {NoStop}%
\bibitem [{\citenamefont {Aalbers}\ \emph {et~al.}(2023{\natexlab{a}})\citenamefont {Aalbers}, \citenamefont {Pelssers}, \citenamefont {Angevaare},\ and\ \citenamefont {Morå}}]{jelle_aalbers_2023_7636982}%
  \BibitemOpen
  \bibfield  {author} {\bibinfo {author} {\bibfnamefont {J.}~\bibnamefont {Aalbers}}, \bibinfo {author} {\bibfnamefont {B.}~\bibnamefont {Pelssers}}, \bibinfo {author} {\bibfnamefont {J.~R.}\ \bibnamefont {Angevaare}},\ and\ \bibinfo {author} {\bibfnamefont {K.~D.}\ \bibnamefont {Morå}},\ }\href {https://doi.org/10.5281/zenodo.7636982} {\bibinfo {title} {Jelleaalbers/wimprates: v0.5.0}} (\bibinfo {year} {2023}{\natexlab{a}})\BibitemShut {NoStop}%
\bibitem [{\citenamefont {Baxter}\ \emph {et~al.}(2021)\citenamefont {Baxter} \emph {et~al.}}]{Baxter:2021pqo}%
  \BibitemOpen
  \bibfield  {author} {\bibinfo {author} {\bibfnamefont {D.}~\bibnamefont {Baxter}} \emph {et~al.},\ }\bibfield  {title} {\bibinfo {title} {{Recommended conventions for reporting results from direct dark matter searches}},\ }\href {https://doi.org/10.1140/epjc/s10052-021-09655-y} {\bibfield  {journal} {\bibinfo  {journal} {Eur. Phys. J. C}\ }\textbf {\bibinfo {volume} {81}},\ \bibinfo {pages} {907} (\bibinfo {year} {2021})},\ \Eprint {https://arxiv.org/abs/2105.00599} {arXiv:2105.00599 [hep-ex]} \BibitemShut {NoStop}%
\bibitem [{\citenamefont {Abdelhameed}\ \emph {et~al.}(2019)\citenamefont {Abdelhameed} \emph {et~al.}}]{CRESST:2019jnq}%
  \BibitemOpen
  \bibfield  {author} {\bibinfo {author} {\bibfnamefont {A.~H.}\ \bibnamefont {Abdelhameed}} \emph {et~al.} (\bibinfo {collaboration} {CRESST}),\ }\bibfield  {title} {\bibinfo {title} {{First results from the CRESST-III low-mass dark matter program}},\ }\href {https://doi.org/10.1103/PhysRevD.100.102002} {\bibfield  {journal} {\bibinfo  {journal} {Phys. Rev. D}\ }\textbf {\bibinfo {volume} {100}},\ \bibinfo {pages} {102002} (\bibinfo {year} {2019})},\ \Eprint {https://arxiv.org/abs/1904.00498} {arXiv:1904.00498 [astro-ph.CO]} \BibitemShut {NoStop}%
\bibitem [{\citenamefont {Angloher}\ \emph {et~al.}(2024)\citenamefont {Angloher} \emph {et~al.}}]{CRESST:2024cpr}%
  \BibitemOpen
  \bibfield  {author} {\bibinfo {author} {\bibfnamefont {G.}~\bibnamefont {Angloher}} \emph {et~al.} (\bibinfo {collaboration} {CRESST}),\ }\bibfield  {title} {\bibinfo {title} {{First observation of single photons in a CRESST detector and new dark matter exclusion limits}},\ }\href {https://doi.org/10.1103/PhysRevD.110.083038} {\bibfield  {journal} {\bibinfo  {journal} {Phys. Rev. D}\ }\textbf {\bibinfo {volume} {110}},\ \bibinfo {pages} {083038} (\bibinfo {year} {2024})},\ \Eprint {https://arxiv.org/abs/2405.06527} {arXiv:2405.06527 [astro-ph.CO]} \BibitemShut {NoStop}%
\bibitem [{\citenamefont {Ressell}\ and\ \citenamefont {Dean}(1997)}]{Ressell:1997kx}%
  \BibitemOpen
  \bibfield  {author} {\bibinfo {author} {\bibfnamefont {M.~T.}\ \bibnamefont {Ressell}}\ and\ \bibinfo {author} {\bibfnamefont {D.~J.}\ \bibnamefont {Dean}},\ }\bibfield  {title} {\bibinfo {title} {{Spin dependent neutralino - nucleus scattering for $A \sim 127$ nuclei}},\ }\href {https://doi.org/10.1103/PhysRevC.56.535} {\bibfield  {journal} {\bibinfo  {journal} {Phys. Rev. C}\ }\textbf {\bibinfo {volume} {56}},\ \bibinfo {pages} {535} (\bibinfo {year} {1997})},\ \Eprint {https://arxiv.org/abs/hep-ph/9702290} {arXiv:hep-ph/9702290} \BibitemShut {NoStop}%
\bibitem [{\citenamefont {O'Hare}(2021)}]{Fog}%
  \BibitemOpen
  \bibfield  {author} {\bibinfo {author} {\bibfnamefont {C.~A.~J.}\ \bibnamefont {O'Hare}},\ }\bibfield  {title} {\bibinfo {title} {New definition of the neutrino floor for direct dark matter searches},\ }\href {https://doi.org/10.1103/PhysRevLett.127.251802} {\bibfield  {journal} {\bibinfo  {journal} {Phys. Rev. Lett.}\ }\textbf {\bibinfo {volume} {127}},\ \bibinfo {pages} {251802} (\bibinfo {year} {2021})}\BibitemShut {NoStop}%
\bibitem [{\citenamefont {Agnes}\ \emph {et~al.}(2018)\citenamefont {Agnes} \emph {et~al.}}]{DarkSide:2018bpj}%
  \BibitemOpen
  \bibfield  {author} {\bibinfo {author} {\bibfnamefont {P.}~\bibnamefont {Agnes}} \emph {et~al.} (\bibinfo {collaboration} {DarkSide}),\ }\bibfield  {title} {\bibinfo {title} {{Low-Mass Dark Matter Search with the DarkSide-50 Experiment}},\ }\href {https://doi.org/10.1103/PhysRevLett.121.081307} {\bibfield  {journal} {\bibinfo  {journal} {Phys. Rev. Lett.}\ }\textbf {\bibinfo {volume} {121}},\ \bibinfo {pages} {081307} (\bibinfo {year} {2018})},\ \Eprint {https://arxiv.org/abs/1802.06994} {arXiv:1802.06994 [astro-ph.HE]} \BibitemShut {NoStop}%
\bibitem [{\citenamefont {Aprile}\ \emph {et~al.}(2019)\citenamefont {Aprile} \emph {et~al.}}]{XENON:2019gfn}%
  \BibitemOpen
  \bibfield  {author} {\bibinfo {author} {\bibfnamefont {E.}~\bibnamefont {Aprile}} \emph {et~al.} (\bibinfo {collaboration} {XENON}),\ }\bibfield  {title} {\bibinfo {title} {{Light Dark Matter Search with Ionization Signals in XENON1T}},\ }\href {https://doi.org/10.1103/PhysRevLett.123.251801} {\bibfield  {journal} {\bibinfo  {journal} {Phys. Rev. Lett.}\ }\textbf {\bibinfo {volume} {123}},\ \bibinfo {pages} {251801} (\bibinfo {year} {2019})},\ \Eprint {https://arxiv.org/abs/1907.11485} {arXiv:1907.11485 [hep-ex]} \BibitemShut {NoStop}%
\bibitem [{\citenamefont {Aalbers}\ \emph {et~al.}(2023{\natexlab{b}})\citenamefont {Aalbers} \emph {et~al.}}]{LZ:2022lsv}%
  \BibitemOpen
  \bibfield  {author} {\bibinfo {author} {\bibfnamefont {J.}~\bibnamefont {Aalbers}} \emph {et~al.} (\bibinfo {collaboration} {LZ}),\ }\bibfield  {title} {\bibinfo {title} {{First Dark Matter Search Results from the LUX-ZEPLIN (LZ) Experiment}},\ }\href {https://doi.org/10.1103/PhysRevLett.131.041002} {\bibfield  {journal} {\bibinfo  {journal} {Phys. Rev. Lett.}\ }\textbf {\bibinfo {volume} {131}},\ \bibinfo {pages} {041002} (\bibinfo {year} {2023}{\natexlab{b}})},\ \Eprint {https://arxiv.org/abs/2207.03764} {arXiv:2207.03764 [hep-ex]} \BibitemShut {NoStop}%
\bibitem [{\citenamefont {Agnese}\ \emph {et~al.}(2018{\natexlab{a}})\citenamefont {Agnese} \emph {et~al.}}]{SuperCDMS:2017mbc}%
  \BibitemOpen
  \bibfield  {author} {\bibinfo {author} {\bibfnamefont {R.}~\bibnamefont {Agnese}} \emph {et~al.} (\bibinfo {collaboration} {SuperCDMS}),\ }\bibfield  {title} {\bibinfo {title} {{Results from the Super Cryogenic Dark Matter Search Experiment at Soudan}},\ }\href {https://doi.org/10.1103/PhysRevLett.120.061802} {\bibfield  {journal} {\bibinfo  {journal} {Phys. Rev. Lett.}\ }\textbf {\bibinfo {volume} {120}},\ \bibinfo {pages} {061802} (\bibinfo {year} {2018}{\natexlab{a}})},\ \Eprint {https://arxiv.org/abs/1708.08869} {arXiv:1708.08869 [hep-ex]} \BibitemShut {NoStop}%
\bibitem [{\citenamefont {Agnese}\ \emph {et~al.}(2014)\citenamefont {Agnese} \emph {et~al.}}]{PhysRevLett.112.041302}%
  \BibitemOpen
  \bibfield  {author} {\bibinfo {author} {\bibfnamefont {R.}~\bibnamefont {Agnese}} \emph {et~al.} (\bibinfo {collaboration} {{SuperCDMS}}),\ }\bibfield  {title} {\bibinfo {title} {Search for low-mass weakly interacting massive particles using voltage-assisted calorimetric ionization detection in the {SuperCDMS} experiment},\ }\href {https://doi.org/10.1103/PhysRevLett.112.041302} {\bibfield  {journal} {\bibinfo  {journal} {Phys. Rev. Lett.}\ }\textbf {\bibinfo {volume} {112}},\ \bibinfo {pages} {041302} (\bibinfo {year} {2014})}\BibitemShut {NoStop}%
\bibitem [{\citenamefont {Aprile}\ \emph {et~al.}(2023)\citenamefont {Aprile} \emph {et~al.}}]{XENON:2023cxc}%
  \BibitemOpen
  \bibfield  {author} {\bibinfo {author} {\bibfnamefont {E.}~\bibnamefont {Aprile}} \emph {et~al.} (\bibinfo {collaboration} {XENON}),\ }\bibfield  {title} {\bibinfo {title} {{First Dark Matter Search with Nuclear Recoils from the XENONnT Experiment}},\ }\href {https://doi.org/10.1103/PhysRevLett.131.041003} {\bibfield  {journal} {\bibinfo  {journal} {Phys. Rev. Lett.}\ }\textbf {\bibinfo {volume} {131}},\ \bibinfo {pages} {041003} (\bibinfo {year} {2023})},\ \Eprint {https://arxiv.org/abs/2303.14729} {arXiv:2303.14729 [hep-ex]} \BibitemShut {NoStop}%
\bibitem [{\citenamefont {Billard}\ \emph {et~al.}(2014{\natexlab{b}})\citenamefont {Billard}, \citenamefont {Strigari},\ and\ \citenamefont {Figueroa-Feliciano}}]{Billard:2013qya}%
  \BibitemOpen
  \bibfield  {author} {\bibinfo {author} {\bibfnamefont {J.}~\bibnamefont {Billard}}, \bibinfo {author} {\bibfnamefont {L.}~\bibnamefont {Strigari}},\ and\ \bibinfo {author} {\bibfnamefont {E.}~\bibnamefont {Figueroa-Feliciano}},\ }\bibfield  {title} {\bibinfo {title} {{Implication of neutrino backgrounds on the reach of next generation dark matter direct detection experiments}},\ }\href {https://doi.org/10.1103/PhysRevD.89.023524} {\bibfield  {journal} {\bibinfo  {journal} {Phys. Rev. D}\ }\textbf {\bibinfo {volume} {89}},\ \bibinfo {pages} {023524} (\bibinfo {year} {2014}{\natexlab{b}})},\ \Eprint {https://arxiv.org/abs/1307.5458} {arXiv:1307.5458 [hep-ph]} \BibitemShut {NoStop}%
\bibitem [{\citenamefont {Meng}\ \emph {et~al.}(2021)\citenamefont {Meng} \emph {et~al.}}]{PandaX_2021}%
  \BibitemOpen
  \bibfield  {author} {\bibinfo {author} {\bibfnamefont {Y.}~\bibnamefont {Meng}} \emph {et~al.} (\bibinfo {collaboration} {PandaX-4T}),\ }\bibfield  {title} {\bibinfo {title} {Dark matter search results from the {PandaX-4T} commissioning run},\ }\href {https://doi.org/10.1103/PhysRevLett.127.261802} {\bibfield  {journal} {\bibinfo  {journal} {Phys. Rev. Lett.}\ }\textbf {\bibinfo {volume} {127}},\ \bibinfo {pages} {261802} (\bibinfo {year} {2021})}\BibitemShut {NoStop}%
\bibitem [{\citenamefont {Agnese}\ \emph {et~al.}(2018{\natexlab{b}})\citenamefont {Agnese} \emph {et~al.}}]{Agnese:2017jvy}%
  \BibitemOpen
  \bibfield  {author} {\bibinfo {author} {\bibfnamefont {R.}~\bibnamefont {Agnese}} \emph {et~al.} (\bibinfo {collaboration} {SuperCDMS}),\ }\bibfield  {title} {\bibinfo {title} {{Low-mass dark matter search with {CDMSlite}}},\ }\href {https://doi.org/10.1103/PhysRevD.97.022002} {\bibfield  {journal} {\bibinfo  {journal} {Phys. Rev.}\ }\textbf {\bibinfo {volume} {D 97}},\ \bibinfo {pages} {022002} (\bibinfo {year} {2018}{\natexlab{b}})},\ \Eprint {https://arxiv.org/abs/1707.01632} {arXiv:1707.01632 [astro-ph.CO]} \BibitemShut {NoStop}%
\bibitem [{\citenamefont {Lee}\ and\ \citenamefont {Weinberg}(1977)}]{Lee-Weinberg}%
  \BibitemOpen
  \bibfield  {author} {\bibinfo {author} {\bibfnamefont {B.~W.}\ \bibnamefont {Lee}}\ and\ \bibinfo {author} {\bibfnamefont {S.}~\bibnamefont {Weinberg}},\ }\bibfield  {title} {\bibinfo {title} {Cosmological lower bound on heavy-neutrino masses},\ }\href {https://doi.org/10.1103/PhysRevLett.39.165} {\bibfield  {journal} {\bibinfo  {journal} {Phys. Rev. Lett.}\ }\textbf {\bibinfo {volume} {39}},\ \bibinfo {pages} {165} (\bibinfo {year} {1977})}\BibitemShut {NoStop}%
\end{thebibliography}%

\end{document}